\PassOptionsToPackage{table, svgnames, dvipsnames}{xcolor}
\documentclass[aps,pra,twocolumn,superscriptaddress,longbibliography,nofootinbib]{revtex4-2}
\usepackage{amsmath, amsfonts, amsthm, amssymb}
\usepackage[margin=1in]{geometry}
\usepackage{graphicx}
\usepackage{tikz}
\usepackage[hidelinks]{hyperref}
\usepackage[caption=false]{subfig}

\usepackage{qcircuit}

\usepackage{dcolumn}
\usepackage{blindtext}
\usepackage{floatrow}
\usepackage[version=3]{mhchem}
\usepackage{footnote}
\usepackage{bm}
\usepackage{dsfont}
\usepackage{xr-hyper}
\usepackage{cleveref}
\usepackage{soul}
\usepackage[normalem]{ulem}

\usepackage{hhline}
\usepackage{booktabs}
\usepackage{algorithm}
\usepackage[noend]{algpseudocode}

\usepackage{colortbl}
\makeatletter
    \def\CT@@do@color{%
      \global\let\CT@do@color\relax
            \@tempdima\wd\z@
            \advance\@tempdima\@tempdimb
            \advance\@tempdima\@tempdimc
    \advance\@tempdimb\tabcolsep
    \advance\@tempdimc\tabcolsep
    \advance\@tempdima2\tabcolsep
            \kern-\@tempdimb
            \leaders\vrule
                    \hskip\@tempdima\@plus  1fill
            \kern-\@tempdimc
            \hskip-\wd\z@ \@plus -1fill }        
\makeatother

\Crefname{equation}{Eq.}{Eqs.}
\Crefname{equation}{Equation}{Equations}
\Crefname{figure}{Fig.}{Figs.} 
\Crefname{figure}{Figure}{Figures}
\Crefname{section}{Sect.}{Sects.}
\Crefname{section}{Section}{Sections}
\Crefname{table}{Table}{Tables}
\Crefname{appsec}{}{Appendices}

\algrenewcommand{\algorithmiccomment}[1]{\hskip3em \texttt{// #1}}

\labelformat{subsection}{\thesection.#1}
\labelformat{subsubsection}{\thesection.\thesubsection.#1}

\newcommand{\torus}{\ensuremath{  
    \begin{tikzpicture}[yscale=cos(60)]
        \draw[double distance=0.75mm] (0:0.15) arc (0:180:0.15);
        \draw[double distance=0.75mm] (180:0.15) arc (180:360:0.15);
  \end{tikzpicture}
  }}%

\newcommand{\cylinder}{\ensuremath{  
    \begin{tikzpicture}
        \draw[fill=white] (0,0) ellipse (0.1 and 0.05);
        \draw (-0.1,0) -- (-0.1,-0.2);
        \draw (-0.1,-0.2) arc (180:360:0.1 and 0.05);
        \draw [dashed,dash pattern=on 1pt off 0.5pt] (-0.1,-0.2) arc (180:360:0.1 and -0.05);
        \draw (0.1,-0.2) -- (0.1,0); 
        \fill [white,opacity=0.5] (-0.1,0) -- (-0.1,-0.2) arc (180:360:0.1 and 0.05) -- (0.1,0) arc (0:180:0.1 and -0.05);
  \end{tikzpicture}
  }}%

\begin{document}

\title{Romanesco codes: Bias-tailored qLDPC codes from fractal codes}

\author{Catherine Leroux}
\email{catilero@amazon.com}
\author{Joseph K. Iverson}
\affiliation{AWS Center for Quantum Computing, Pasadena, CA 91125, USA}

\begin{abstract}
We introduce 
and analyze a family of Clifford-deformed bivariate bicycle codes that are tailored for biased noise. Our qLDPC codes are defined on a bipartite hexagonal lattice with limited-range gates and low-weight stabilizers. The code is non-CSS, featuring stabilizer generators that are each half X and half Z. We find small examples with high encoding rate that perform well for a large range of bias. In the limit of large noise bias, the code reduces to two independent classical cellular automaton codes, giving a distance scaling better than is possible with 2D topological quantum codes. Our construction combines two classical cellular automaton codes, LDPC codes that were recently proposed for use with noise-biased cat qubits, related to each other by a reflection. Each stabilizer in the quantum code is obtained by multiplying an all-X stabilizer from the first code with an all-Z stabilizer from the second code, or the other way around. The result is a self-dual quantum code with a number of qubits equal to the sum of the input codes and stabilizer weight and locality determined by the input codes. Under strong noise bias, the effective distance of the quantum code approaches the distance of the input codes. We simulate the logical performance of our qLDPC codes under code-capacity noise and find strong suppression of the logical error rate.
\end{abstract}

\maketitle

\begin{figure*}[t!]
    \centering
    \includegraphics[width=\linewidth]{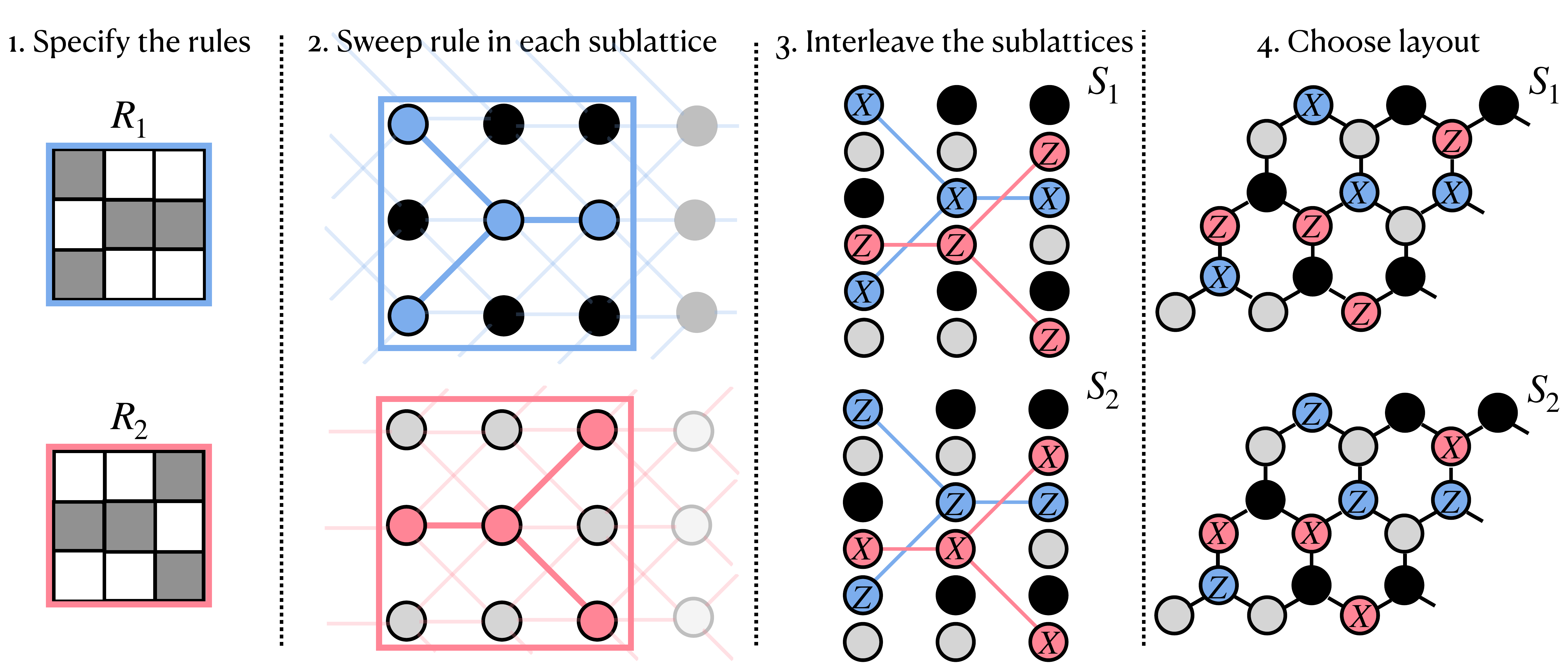}
    \caption{Schematic representation of our code construction. 1. We specify the stabilizer shape of the cellular automaton code on each sub-lattice, black or gray, as a binary matrix labelled $R_1$ or $R_2$. Here $R_2$ is related to $R_1$ by a $180^\circ$ rotation so that all stabilizers commute. 2. We build the parity check matrices of the two cellular automaton codes by sweeping the binary matrices $R_1$ and $R_2$ across the sub-lattices, giving us all the stabilizers of the classical codes. Qubits that are part of the same stabilizer in either the black or gray sub-lattice are highlighted in blue or red and connected by edges for clarity. 3. We create the stabilizers $S_1$, $S_2$ of the quantum code by interleaving the sub-lattices of the two classical codes. 4. We choose a bipartite honeycomb lattice layout for the quantum code motivated by locality in the hardware device.
    }
    \label{fig:construction}
\end{figure*}

\section{Introduction}\label{sec:introduction}

Past works have suggested that biased-noise qubits, such as cat qubits~\cite{Cochrane1999,Lescanne2020,putterman2024}, can reduce the number of resources required to realize quantum error correction with strong error suppression~\cite{Aliferis2008,Aliferis2009,Tuckett2018,Chamberland2022}. In the limit of large noise bias, where the qubit effectively acts like a classical bit, an attractive option is to use classical error correcting codes such as the repetition code~\cite{Regent2023,Putterman2025,Gouzien2023,Guillaud2019,Guillaud2021}. A concatenated cat-repetition code has already been experimentally demonstrated~\cite{Putterman2025}. However, the repetition code encodes only a single bit. Recently, the use of a specific family of classical LDPC codes, known as cellular automaton codes, was proposed to correct phase-flip errors in strongly noise-biased cat qubits while maintaining some locality in the checks~\cite{Ruiz2025}. Although the distance and encoding rate of these codes are excellent relative to repetition codes, these are classical codes that offer no protection against bit-flip errors.

At finite noise-bias, bit-flip errors can no longer be ignored for sufficiently long fault-tolerant computations. Past works have suggested that a rectangular surface code, with greater distance against either $X$ or $Z$ errors, could be used to correct both types of errors~\cite{hann2024} while taking advantage of bias in the noise. It was also shown that the performance of topological codes under biased noise can be improved via local Clifford deformations. The XZZX code~\cite{Bonilla2021}, a bias-tailored topological code resulting from local Hadamard rotations on half of the data qubits in the surface code, was shown to have a large threshold under code-capacity noise~\cite{Darmawan2021}. Other bias-tailored 2D topological codes have been proposed such as the  XY code~\cite{Tuckett2019}, the cyclic XYZ or XYZ$^2$ code~\cite{Srivastava2022,liang2025}, generalized XZZX codes~\cite{xu2022}, the XYZ code~\cite{Miguel2023}, Clifford-deformed compass codes~\cite{campos2024}, randomly Clifford-deformed surface codes~\cite{Dua2024} and bias-tailored Floquet codes~\cite{setiawan2024}. These codes all promise improved performance under biased-noise, but low encoding rates.

In recent years, qLDPC codes~\cite{Kovalev2012,Tillich2014,Breuckmann2021_qldpc,Breuckmann2021_balanced,Hastings2021,Panteleev2022_linear,Panteleev2022_good} have attracted a lot of interest since they achieve higher encoding rates than topological codes. Although these codes can have asymptotically ``good" code parameters~\cite{Panteleev2022_good}, they require long-range gates and generally lack a natural embedding in a 2D plane, posing a challenge to experimental implementation, particularly with superconducting qubits. The bivariate bicycle codes~\cite{kovalev2013quantum, Bravyi2024} are lifted product codes with toric embeddings that have emerged as potentially promising candidates for experimental realization because of their planar layouts, weight-6 stabilizers, and small examples with high encoding rate and logical performance. Similarly to topological codes, it is possible to use local Clifford deformations to tailor qLDPC codes to noise bias. Indeed, a scheme to construct bias-tailored lifted product codes has been recently proposed~\cite{Roffe2023}. This scheme relies on applying local Hadamard rotations on half of the data qubits of a lifted product code, similarly to the XZZX code, conserving the code parameters of the original CSS code. Under this transformation, the lifted product code example in \cite{Roffe2023} shows logical error rate improvement from biased noise that saturates at quite modest values of bias.

In this work, we introduce and analyze a family of self-dual bivariate bicycle codes, whose input classical codes are cellular automaton codes. We apply Hadamard rotations on half of the data qubits in order to tailor these codes to biased noise, similarly to other Clifford-deformed codes~\cite{Tuckett2019,Bonilla2021,Darmawan2021,Roffe2023,Miguel2023,Srivastava2022,liang2025,xu2022,campos2024,Dua2024,setiawan2024}. In the infinite noise bias limit, where Z errors dominate, the decoding graph decouples into two sub-lattices, where each sub-lattice supports one of the two input classical codes. Under strongly biased noise, the effective distance~\cite{xu2022,Dua2024} of the code then approaches the distance of the input classical codes. We numerically search through automaton rules and find small qLDPC codes offering high encoding rate, low-weight and translation-invariant stabilizer generators and a toric layout. In addition, we observe strong improvement in logical error rate beginning with experimentally accessible bias and continuing into the large bias regime. We study codes on the torus but we also show examples with open boundary conditions. We refer to our codes as the \emph{Romanesco} codes, due to the fractal nature of the Romanesco broccoli.

\section{Preliminaries} \label{sec: Preliminaries}

\subsection{Bivariate bicycle codes}

Bivariate bicycle codes are qLDPC codes defined on a torus of dimension $l\times m$ by a pair of matrices $A$ and $B$, such that the X and Z parity check matrices are $H_X = [A | B ]$ and $H_Z = [B^T | A^T]$~\cite{Bravyi2024}. There are $n=2lm$ data qubits. $A$ and $B$ can be expressed as bivariate polynomials of square cyclic matrices with dimensions equal to those of the torus. Defining the two cyclic variables $x = S_l \otimes I_m$ and $y = I_l \otimes S_m$ where $[S_l]_{ij}=\delta_{j,(i+1)\phantom{.}\mathrm{mod}\phantom{.}l}$ is the $l\times l$ cyclic shift matrix and $I_l$ is the $l\times l$ identity matrix, we write for example $A = \sum_{k=1}^w x^{p_k} y^{q_k}$ where $w$ is the weight of the stabilizers in $A$ and $q_k$ ($p_k$) is the degree of the $x$ ($y$) variable in the $k$th monomial of $A$. The stabilizers of the quantum code are guaranteed to commute by construction: $H_X \cdot H_Z^T = 0$ in mod 2 space since $[x,y]=[A, B] =0$.

\subsection{Noise bias}

We consider a Pauli noise model where each data qubit has X, Y, and Z Pauli error rates $p_X$, $p_Y$ and $p_Z$, respectively. We define the noise bias $\eta$ through the relation $p_X = p_Y = p_Z/\eta$.

In the case of cat qubits, phase-flips dominate over bit-flips ($p_Z \gg p_X, p_Y$). This asymmetry results from the two pointer states of the cat qubit being large coherent states of opposite phase: the probability of a bit-flip is proportional to the overlap of their wavefunctions which is exponentially suppressed with respect to the distance between them~\cite{Cochrane1999,Lescanne2020,putterman2024}. This distance can be tuned in situ and made arbitrarily large. While the bit-flip rate is exponentially suppressed in the photon number, the phase-flip rate grows linearly with it. This gives a tunably-biased noise model.

\subsection{Bias-tailored color code} \label{subsec: XZ$^3$ code}

The Romanesco code family that we explore in this work includes the Clifford-deformed color code. Particularly, here we focus on the honeycomb color code with parameters $[[18q^2, 4, 4q]]$ introduced in~\cite{deCarvalho2021}, which is a CSS code defined on a square region with periodic boundary conditions, and where $q$ is the number of times each color repeats along each boundary (i.e. there are $3q$ hexagons along each boundary). This code has two weight-6 stabilizers per plaquette, labeled $S_1$ and $S_2$, where $S_1$ is purely X-type while $S_2$ is Z-type (see \cref{app:color code to fractal codes} for illustrations). The lattice is bipartite, and we label the two triangular sub-lattices black and gray. The honeycomb color code is a bivariate bicycle code where the matrices $A$ and $B\equiv A^T$ specify the structure of the X/Z checks on the black and gray sub-lattices, respectively.

We define the XZ$^3$ color code as the non-CSS code obtained by applying a Hadamard gate, mapping $X\to Z$ and $Z\to X$, on each gray qubit of the honeycomb color code similarly to what was done for the XYZ color code~\cite{Miguel2023}. The code parameters are conserved under this unitary rotation. However, the logical performance of the CSS and non-CSS codes will differ in the presence of biased noise. 

In the large-bias limit ($p_Z \gg p_X, p_Y$), XZ$^3$ code stabilizers reduce to weight-3 X-type stabilizers on each sub-lattice. The parity check matrices on the black and gray sub-lattices have identical structure given by a classical cellular automaton code~\cite{Miguel2023}. Therefore, the all-Z logical representatives with lowest weight for the code are logical representatives of the classical cellular automaton codes supported on only one of the sub-lattices. It follows that the effective code distance in the infinite bias regime is the classical code distance.

\subsection{Cellular Automaton Codes}

The stabilizer generators of a cellular automaton or fractal code are specified by a binary matrix $R$ where the top row has a single non-zero entry. $R$ also specifies a cellular automaton rule for the code as follows: the value of the bit in the top row is set to the parity of all non-zero elements in the lower rows. The stabilizer generators of the code are translation-invariant parity check operators acting on the non-zero elements of $R$. These codes have been shown to yield a good trade-off between the code distance and the encoding rate while having quasi-local checks~\cite{Ruiz2025}.

In this work, we introduce a family of Clifford-deformed bivariate bicycle codes, called \emph{Romanesco} codes, that uses two cellular automaton codes as input codes and include the XZ$^3$ color code.

\section{Romanesco Codes} \label{sec: Romanesco codes}

\subsection{Code construction} \label{subsec: code construction}

The Romanesco codes $C_\clubsuit$ are obtained by applying a Clifford deformation on a family of bivariate bicycle codes $C$, which are self-dual (i.e. $H_X = H_Z$) like the color code, and whose input codes are built from cellular automaton codes.

\emph{CSS code construction--} Let $R_1$ and $R_2$ be the translation-invariant stabilizer generators of the two input cellular automaton codes, respectively. We construct the polynomials $A$ and $B$ of the bivariate bicycle code $C$ from $R_1$ and $R_2$, by reading off the degrees of the monomials directly from the non-zero entries of the matrices $R_1$ and $R_2$, e.g. $[R_1]_{ij}=1$ becomes the term $x^i y^j$ in A. In order to have a self-dual code, where there is an X- and a Z-type generator for each plaquette, we constrain $R_2$ to be a $180^\circ$ rotation of $R_1$ so that $B\equiv A^T$. $C$ therefore admits transversal logical H, S and CNOT gates similarly to the two-dimensional color code~\cite{Kubica2015}.

A geometric way of understanding our code construction is shown in \cref{fig:construction} in the case where we lay out each classical code on one of the two sub-lattices of a bipartite honeycomb lattice so that $R_1$ acts on black qubits whereas $R_2$ acts on gray qubits. For each plaquette we apply $R_1$ on the black qubits with the bottom left black qubit of the plaquette labeled with the bottom left element of $R_1$, and similarly apply $R_2$ on the gray qubits. The non-zero elements of $R_1$ and $R_2$ give us the data qubits involved in the two stabilizers $S_1$ and $S_2$ acting on the plaquette.

\emph{Clifford deformation--} We partition the parity check matrices $H_X$ and $H_Z$ in two sectors: the first half of the columns are applied to black qubits and the second half to gray qubits. We transform $C$ into a non-CSS Romanesco code $C_\clubsuit$ by applying a local Hadamard rotation on each gray qubit (in the second sector of the parity check matrices). The full parity check matrix transforms as
\begin{equation}
\label{eq:HadamardRotation}
    \left(
    \begin{array} {c c | c c}
        H_X^b & H_X^g & 0 & 0 \\
        0 & 0 & H_Z^b & H_Z^g
    \end{array}
    \right)
    \to
    \left(
    \begin{array} {c c | c c}
        H_X^b & 0 & 0 & H_Z^g \\
        0 & H_X^g & H_Z^b & 0
    \end{array}
    \right),
\end{equation}
where $H_{X/Z}^b$ is the first sector of $H_{X/Z}$ and $H_{X/Z}^g$ is the second sector. These match the matrices defined above, $H_X^b = A$ and $H_X^g = B = A^T$. The X and Z stabilizers of $C$ now become mixed stabilizers alternating between X and Z checks in $C_\clubsuit$, like in the XZ$^3$ code. This Clifford deformation on qLDPC codes was first proposed in \cite{Roffe2023}, however here we use fractal codes for the input classical codes.

\emph{Code parameters--} We denote the code parameters of $C_\clubsuit$ (which are identical to $C$) as $[[n,k,d]]$ and the code parameters of the input cellular automaton codes as $[n_c,k_c,d_c]$ where $n_c = n/2$. Here $d\leq d_c$ and $k = n - \mathrm{rk}(H_X) - \mathrm{rk}(H_Z) \leq 2k_c$ where $k_c = n/2 - \mathrm{rk}(A)$ and $\mathrm{rk}$ is the rank mod 2.

The Pauli-X parts of the stabilizer generators in~\cref{eq:HadamardRotation} are supported on separate sectors of the qubits. Therefore, in the infinite bias limit where the Z-type checks can be ignored, the decoding graph has at least two decoupled sub-lattices. Our codes exactly reduce to two copies of cellular automaton codes which can have large distance since they are each supported on half of the data qubits. In comparison, The XZZX code reduces to several copies of the repetition code with a smaller distance unless twisted, periodic boundary conditions are used~\cite{Roffe2023,xu2022}.

\subsection{Numerical search}

We numerically search to identify the most promising codes (see \cref{app:Numerical search} for detailed explanations). We first generate all unique stabilizer generators $R$ of size $m\times m$ with $m\in{2,3,4}$ and with weight $w \in {3, 4}$ (such that we have weight-$w$ stabilizers in the classical code and therefore weight-$2w$ stabilizers in the quantum code). We consider periodic square lattices for the cellular automaton codes with $9 q^2$ qubits (such that the quantum codes have $18 q^2$ qubits) with $1\leq q \leq 5$. We use this geometric constraint to find the most promising stabilizer shapes. For each shape we identify code families with promising code parameters by generating instances on rectangular lattices.

The best code families we find for moderate size lattices are shown in \cref{tab:Romanesco codes rectangular}. Here we define the families by keeping the number of encoded qubits $k$ fixed (see \cref{app:Numerical search} for further details). The scalings for $k$, $d$ and $d_c$ are found empirically by generating multiple lattice sizes. Some families have a quantum distance that is equal to the minimum dimension of the lattice suggesting that they perform better on square lattices while others have a quantum distance that is proportional to the width and are therefore better suited on rectangular lattices. 

In the infinite bias limit, the lowest-weight all-Z logical operators have length $d_c$, and therefore we expect the effective distance of the code to be $d_c$. Importantly, for all of the families in \cref{tab:Romanesco codes rectangular}, $d_c$ scales linearly with the number of data qubits $n$ in the code. We define the overhead reduction factor $v_\infty = k d_\infty/n$~\cite{Ruiz2025}, where $d_\infty$ is the code distance at infinite noise bias ($d_\infty = d_c$ in our construction). This ratio captures the overhead of the code relative the repetition code, which satisfies $v_\infty = 1$. The twisted XZZX code~\cite{Roffe2023,xu2022} also satisfies $v_\infty = 1$. As shown in \cref{tab:Romanesco codes rectangular} our code construction has $v_\infty > 1$ in some cases showing that our codes can achieve a better trade-off between having a high encoding rate and a large infinite bias distance. All code families in \cref{tab:Romanesco codes rectangular} are defined on two-dimensional Euclidean lattices with local connectivity and therefore satisfy the Bravyi-Poulin-Terhal (BPT) bound $kd^2 = \mathcal{O}(n)$ and  the  Bravyi-Terhal (BT) bound $d = \mathcal{O}(\sqrt{n})$~\cite{delfosse2021,Baspin2022,Baspin2023}. However $kd_c = \mathcal{O}(n)$ and therefore in the large bias limit, where the effective code distance is approximately $d_c$, our codes could in principle achieve a linear scaling.

\begin{table}[t!]
    \centering
    \caption{Families of Romanesco codes on a bipartite honeycomb lattice, where each sub-lattice lies on a torus with height $H$ and width $L$ so that there are $N=2 HL$ data qubits in the quantum codes. We also define the quantity $D_{\mu,\nu} = \min(\mu H, \nu L)$. Here we report the code families with the highest encoding rates and $v_\infty = kd_c > 1$ per stabilizer shape. We did not find any code family with better parameters for $m=4$ than for $m=3$ with a significant number of examples (with less than 1000 qubits) to get reliable scalings. The code families that are highlighted have $v_\infty = kd_c/n> 2$ and are considered the best families.}
    \label{tab:Romanesco codes rectangular}
    \begin{tabular}{c|c|c|c|c|c}
        %\hline
        % \rowcolor{lightgray}
        \cellcolor{black!10}{$\bm{[[n,k,d]]}_\clubsuit^{\torus}$} & \cellcolor{black!10}{$\bm{d_c}$} & \cellcolor{black!10}{$\bm{v_\infty}$} & \cellcolor{black!10}{\textbf{Constraints}} & \cellcolor{black!10}{$\bm{R_1}$} & \cellcolor{black!10}{$\bm{R_2}$}\\ \hline

        $[[N,4,D_{4/3,4/3}]]$ & 
        $\dfrac{N}{3}$ & 1.33 &
        $\substack{H\%3 =0, \ H\geq 3 \\ L\% 3 = 0, \ L\geq 3 \\ H \% 6 \neq 0  \text{ or }  L \% 6 \neq 0}$ & 
        \begin{tikzpicture}[baseline=0.5em]
        \draw[white, very thick] (0,0.6) rectangle (0,0.7);
        \draw[black, very thick, fill=lightgray] (0,0) rectangle (0.2,0.2);
        \draw[black, very thick, fill=lightgray] (0.2,0) rectangle (0.4,0.2);
        \draw[black, very thick, fill=lightgray] (0,0.2) rectangle (0.2,0.4);
        \draw[white, very thick] (0,-0.1) rectangle (0,0.0);
        \end{tikzpicture} &
        \begin{tikzpicture}[baseline=0.5em]
        \draw[white, very thick] (0,0.4) rectangle (0,0.5);
        \draw[black, very thick, fill=lightgray] (0.2,0) rectangle (0.4,0.2);
        \draw[black, very thick, fill=lightgray] (0,0.2) rectangle (0.2,0.4);
        \draw[black, very thick, fill=lightgray] (0.2,0.2) rectangle (0.4,0.4);
        \draw[white, very thick] (0,-0.1) rectangle (0,0.0);
        \end{tikzpicture}\\ \hline
        
        \cellcolor{orange!0}{$[[N,8,D_{1,1}]]$} & 
        \cellcolor{orange!0}{$\dfrac{N}{4}$} & 2.00 &
        \cellcolor{orange!0}{$\substack{(H-2)\%4=0, \ H \geq 6 \\ L\%4=0, \ L\geq 4 \\ (H, L) \leftrightarrow (L, H)}$} & 
        \cellcolor{orange!0}{\begin{tikzpicture}[baseline=0.5em]
        \draw[orange!0, very thick] (0,0.6) rectangle (0,0.7);
        \draw[black, very thick, fill=lightgray] (0.2,0) rectangle (0.4,0.2);
        \draw[black, very thick, fill=lightgray] (0.2,0.2) rectangle (0.4,0.4);
        \draw[black, very thick, fill=lightgray] (0.4,0.2) rectangle (0.6,0.4);
        \draw[black, very thick, fill=lightgray] (0,0.4) rectangle (0.2,0.6);
        \draw[orange!0, very thick] (0,-0.1) rectangle (0,0.0);
        \end{tikzpicture}} &
        \cellcolor{orange!0}{
        \begin{tikzpicture}[baseline=0.5em]
        \draw[orange!0, very thick] (0,0.6) rectangle (0,0.7);
        \draw[black, very thick, fill=lightgray] (0.4,0) rectangle (0.6,0.2);
        \draw[black, very thick, fill=lightgray] (0,0.2) rectangle (0.2,0.4);
        \draw[black, very thick, fill=lightgray] (0.2,0.2) rectangle (0.4,0.4);
        \draw[black, very thick, fill=lightgray] (0.2,0.4) rectangle (0.4,0.6);
        \draw[orange!0, very thick] (0,-0.1) rectangle (0,0.0);
        \end{tikzpicture}}\\ \hline

        \cellcolor{orange!10}{$[[N,12,D_{1,1}]]$} & 
        \cellcolor{orange!10}{$\dfrac{3N}{16}$} & \cellcolor{orange!10}{2.25} &
        \cellcolor{orange!10}{$\substack{H\%4=0, \ H \geq 4 \\ L\%4=0, \ L \geq 4 \\ H \% 8 \neq 0 \text{ or } L \% 8 \neq 0}$} & 
        \cellcolor{orange!10}{\begin{tikzpicture}[baseline=0.5em]
        \draw[orange!10, very thick] (0,0.6) rectangle (0,0.7);
        \draw[black, very thick, fill=lightgray] (0.2,0) rectangle (0.4,0.2);
        \draw[black, very thick, fill=lightgray] (0.2,0.2) rectangle (0.4,0.4);
        \draw[black, very thick, fill=lightgray] (0.4,0.2) rectangle (0.6,0.4);
        \draw[black, very thick, fill=lightgray] (0,0.4) rectangle (0.2,0.6);
        \draw[orange!10, very thick] (0,-0.1) rectangle (0,0.0);
        \end{tikzpicture}} &
        \cellcolor{orange!10}{
        \begin{tikzpicture}[baseline=0.5em]
        \draw[orange!10, very thick] (0,0.6) rectangle (0,0.7);
        \draw[black, very thick, fill=lightgray] (0.4,0) rectangle (0.6,0.2);
        \draw[black, very thick, fill=lightgray] (0,0.2) rectangle (0.2,0.4);
        \draw[black, very thick, fill=lightgray] (0.2,0.2) rectangle (0.4,0.4);
        \draw[black, very thick, fill=lightgray] (0.2,0.4) rectangle (0.4,0.6);
        \draw[orange!10, very thick] (0,-0.1) rectangle (0,0.0);
        \end{tikzpicture}}\\ \hline
        
        \cellcolor{orange!0}{$[[N, 12, D_{1,2/3}]]$} & 
        \cellcolor{orange!0}{$\dfrac{N}{6}$} & 2.00 &
        \cellcolor{orange!0}{$\substack{H\% 6 =0, \ H\geq 6 \\ L\%3=0, \ L\geq 6 \\ H \% 12 \neq 0}$} & 
        \cellcolor{orange!0}{\begin{tikzpicture}[baseline=0.5em]
        \draw[orange!0, very thick] (0,0.6) rectangle (0,0.7);
        \draw[black, very thick, fill=lightgray] (0,0) rectangle (0.2,0.2);
        \draw[black, very thick, fill=lightgray] (0.2,0.2) rectangle (0.4,0.4);
        \draw[black, very thick, fill=lightgray] (0.4,0.2) rectangle (0.6,0.4);
        \draw[black, very thick, fill=lightgray] (0,0.4) rectangle (0.2,0.6);
        \draw[orange!0, very thick] (0,-0.1) rectangle (0,0.0);
        \end{tikzpicture}} &
        \cellcolor{orange!0}{\begin{tikzpicture}[baseline=0.5em]
        \draw[orange!10, very thick] (0,0.6) rectangle (0,0.7);
        \draw[black, very thick, fill=lightgray] (0.4,0) rectangle (0.6,0.2);
        \draw[black, very thick, fill=lightgray] (0,0.2) rectangle (0.2,0.4);
        \draw[black, very thick, fill=lightgray] (0.2,0.2) rectangle (0.4,0.4);
        \draw[black, very thick, fill=lightgray] (0.4,0.4) rectangle (0.6,0.6);
        \draw[orange!0, very thick] (0,-0.1) rectangle (0,0.0);
        \end{tikzpicture}}\\ \hline
        
        \cellcolor{orange!0}{$[[N, 12, D_{4/3,2/3}]]$} & 
        \cellcolor{orange!0}{$\dfrac{N}{6}$} & 2.00 &
        \cellcolor{orange!0}{$\substack{H\% 12 =0, \ H\geq 12 \\ L\%3=0, \ L\geq 9 \\ L \% 6 \neq 0 }$} & 
        \cellcolor{orange!0}{\begin{tikzpicture}[baseline=0.5em]
        \draw[orange!0, very thick] (0,0.6) rectangle (0,0.7);
        \draw[black, very thick, fill=lightgray] (0,0) rectangle (0.2,0.2);
        \draw[black, very thick, fill=lightgray] (0.2,0.2) rectangle (0.4,0.4);
        \draw[black, very thick, fill=lightgray] (0.4,0.2) rectangle (0.6,0.4);
        \draw[black, very thick, fill=lightgray] (0,0.4) rectangle (0.2,0.6);
        \draw[orange!0, very thick] (0,-0.1) rectangle (0,0.0);
        \end{tikzpicture}} &
        \cellcolor{orange!0}{\begin{tikzpicture}[baseline=0.5em]
        \draw[orange!10, very thick] (0,0.6) rectangle (0,0.7);
        \draw[black, very thick, fill=lightgray] (0.4,0) rectangle (0.6,0.2);
        \draw[black, very thick, fill=lightgray] (0,0.2) rectangle (0.2,0.4);
        \draw[black, very thick, fill=lightgray] (0.2,0.2) rectangle (0.4,0.4);
        \draw[black, very thick, fill=lightgray] (0.4,0.4) rectangle (0.6,0.6);
        \draw[orange!0, very thick] (0,-0.1) rectangle (0,0.0);
        \end{tikzpicture}}\\ \hline

        \cellcolor{orange!10}{$[[N, 16, D_{1,2/3}]]$} & 
        \cellcolor{orange!10}{$\dfrac{N}{6}$} & \cellcolor{orange!10}{2.67} &
        \cellcolor{orange!10}{$\substack{H\% 12 =0, \ H\geq 12 \\ L\%6=0, \ L\geq 6 \\ L\% 12 \neq 0}$} & 
        \cellcolor{orange!10}{\begin{tikzpicture}[baseline=0.5em]
        \draw[orange!10, very thick] (0,0.6) rectangle (0,0.7);
        \draw[black, very thick, fill=lightgray] (0,0) rectangle (0.2,0.2);
        \draw[black, very thick, fill=lightgray] (0.2,0.2) rectangle (0.4,0.4);
        \draw[black, very thick, fill=lightgray] (0.4,0.2) rectangle (0.6,0.4);
        \draw[black, very thick, fill=lightgray] (0,0.4) rectangle (0.2,0.6);
        \draw[orange!10, very thick] (0,-0.1) rectangle (0,0.0);
        \end{tikzpicture}} &
        \cellcolor{orange!10}{\begin{tikzpicture}[baseline=0.5em]
        \draw[orange!10, very thick] (0,0.6) rectangle (0,0.7);
        \draw[black, very thick, fill=lightgray] (0.4,0) rectangle (0.6,0.2);
        \draw[black, very thick, fill=lightgray] (0,0.2) rectangle (0.2,0.4);
        \draw[black, very thick, fill=lightgray] (0.2,0.2) rectangle (0.4,0.4);
        \draw[black, very thick, fill=lightgray] (0.4,0.4) rectangle (0.6,0.6);
        \draw[orange!10, very thick] (0,-0.1) rectangle (0,0.0);
        \end{tikzpicture}}\\ \hline
        
    \end{tabular}
\end{table}

\subsection{Effective distance}\label{subsec:logicals}

The distance of the classical codes $d_c$ is a good indicator of the performance of the quantum code in the large bias limit; however, it is unclear how good the code is at finite noise bias. The notion of an effective distance $d^\prime$ in the presence of noise bias has been introduced before~\cite{xu2022,Dua2024} to capture the scaling of logical error rate for different values of noise strength and bias (logical error rate $p_L \sim p_Z^{\lfloor d^\prime/2 \rfloor}$). Depending on $p_X$ and $p_Z$, the logical error rate will be dominated by logical representatives with different numbers of $X$ and $Z$ Pauli operators. A logical representative with Hamming weight $L(s)$ with $s$ Pauli X operators has a modified weight $w = \log (p_X^s p_Z^{L(s) -s}) /\log(p_Z) = L(s) + s\delta$ where $\delta=-\log(\eta)/\log(p_Z)$. We define the effective distance $d^\prime$ as the minimum modified weight of the logical operators.

The logical operators of the Romanesco codes with all Pauli Z operators are the logical operators of the two classical codes of distance $d_c$ ($L(0)=d_c$). Therefore, the effective distance is approximately $d_c$ for $\delta \gg 0$ but drops to $d$, the quantum distance of the Romanesco code, as $\delta\to 0$. The quantum distance is found at $s=d/2$, where $L(d/2)=d$. These logical operators are composed of half X and half Z because all the lowest-weight all-Z logical representatives with Hamming weight $d$ in the CSS code have an equal number of black and gray qubits, and in the non-CSS code the Pauli operators on the gray qubits are mapped to X operators. Then, a Romanesco code of distance $d$ can tolerate $\lfloor d/4 \rfloor$ bit-flip errors. We therefore expect the logical error rate to be suppressed by $\eta^{-\lfloor d/4 \rfloor}$ when $\delta \to 0$.

In addition to the bias, the effective distance $d^\prime$ depends on the physical error rate $p_Z$. The value of $\delta$ goes to zero as $p_Z$ goes to zero. In particular, although the effective distance is approximately $d_c$ for large $\eta$, we expect that $d^\prime \to d$ at low physical error rates, so that the exponent in the logical error rate $p_L$ decreases as $p_Z$ is lowered.

We can gain more understanding of the effective distance by generating the lowest-weight logical representatives with mixed X and Z operators. We generate $L(s)$ for $0\leq s \leq d/2$ by solving a modified decoding problem where the parity check matrix is augmented with one random test logical at a time~\cite{Bravyi2024} under biased noise (see \cref{app:logical representatives} for further explanations). Since this algorithm works for CSS codes only, as it relies on treating $H_X$ and $H_Z$ independently, we use the parity check matrices of the CSS version of the Romanesco code and Hadamard rotate the noise channels on the gray qubits (see \cref{app:rotating decoding problem}). This approach is heuristic and relies on the BP+OSD-CS decoder~\cite{Roffe2020,Roffe_LDPC_Python_tools_2022}, so we generate many samples with different values for $p_Z$ and $p_X = p_Y = p_Z/\eta$, where $\eta = p_Z/p_X$ is the noise bias, to generate a list of low-weight logical representatives for the (non-CSS) Romanesco code. We then count the number of X operators in each representative. For each number $s$ of X Pauli operators we record the smallest distance $L(s)$ from all the logical representative that we found. The distances for every $s$ value with even lattice dimensions and for each code family identified in \cref{tab:Romanesco codes rectangular} are shown in \cref{tab:mixed logicals families}. For $H,L\geq12$, we find that the $s=2$ representatives have a weight $L(2)\simeq 2L(0)/3$ that grows linearly with the number of data qubits, and we designate these, fractal-like. For smaller $H,L$ or larger $s$ values, the representatives are string-like, meaning that their weight is set by the smallest lattice dimension. We expect that as we increase the lattice dimensions, mixed logical representatives become more and more fractal-like for larger $s$ values.

\begin{table}[t!]
    \centering
    \caption{Distance $d_s$ of the lowest-weight logical representative with $s$ non-Z Pauli operators for the periodic code families in \cref{tab:Romanesco codes rectangular}. Here $D_{\mu,\nu}=\min(\mu H, \nu L)$ denotes the quantum code distance. For the family with $D_{1, 1}$, setting $H=L$ maximizes quantum distance for a given $N$. Increasing $L$ while keeping $H$ fixed does not increase the quantum distance but does increase the infinite bias distance, as shown in the column with $s = 0$. We focus on even dimensions $H\%2 = L\%2 = 0$ for simplicity. ${}^*$\textit{These scalings are approximate and slightly vary depending on the lattice dimensions.} See \cref{app:logical representatives} for additional examples with square lattices.}
    \label{tab:mixed logicals families}
    \begin{tabular}{c c c c c c c c}
        %\hline
        & \multicolumn{7}{c}{\cellcolor{black!10}{$\bm{s}$}} \\
        %\hline
        & \cellcolor{black!10}{$\bm{0}$} & \cellcolor{black!10}{$\bm{1}$} & \cellcolor{black!10}{$\bm{2}$} & \cellcolor{black!10}{$\bm{3}$} & \cellcolor{black!10}{$\bm{4}$} & \cellcolor{black!10}{$\bm{5}$} & \cellcolor{black!10}{$\bm{6}$} \\ %\hline

        \cellcolor{black!10}{$\bm{D_{1,1}}$} & \multicolumn{7}{c}{\cellcolor{black!10}{$[[N, 12, D_{1,1}]]$}} \\ %\hline

        \cellcolor{black!10}{$4$} & \cellcolor{orange!10}{$3N/16$} & \cellcolor{orange!10}{$\varnothing$} & \cellcolor{orange!60}{$4$} & \cellcolor{black!75} & \cellcolor{black!75} & \cellcolor{black!75}  & \cellcolor{black!75} \\ %\hline
        
        \cellcolor{black!10}{$8$} & \cellcolor{orange!10}{$3N/16$} & \cellcolor{orange!10}{$\varnothing$} & \cellcolor{orange!60}{$28$} & \cellcolor{orange!60}{$\varnothing$} & \cellcolor{orange!90}{8} & \cellcolor{black!75}  & \cellcolor{black!75} \\ %\hline

        \cellcolor{black!10}{$12$} & \cellcolor{orange!10}{$3N/16$} & \cellcolor{orange!10}{$\varnothing$} & \cellcolor{orange!30}{${N/8}^*$} & \cellcolor{orange!30}{$\varnothing$} & \cellcolor{orange!60}{24} & \cellcolor{orange!60}{$\varnothing$} & \cellcolor{orange!90}{$12$} \\ %\hline

        \cellcolor{black!10}{$\bm{D_{1,2/3}}$} & \multicolumn{7}{c}{\cellcolor{black!10}{$[[N, 12, D_{1,2/3}]]$}} \\ %\hline

        \cellcolor{black!10}{$4$} & \cellcolor{orange!10}{$N/6$} & \cellcolor{orange!10}{$\varnothing$} & \cellcolor{orange!90}{4} & \cellcolor{black!75} & \cellcolor{black!75} & \cellcolor{black!75} & \cellcolor{black!75}\\ %\hline

        \cellcolor{black!10}{$6$} & \cellcolor{orange!10}{$N/6$} & \cellcolor{orange!10}{$\varnothing$} & \cellcolor{orange!60}{12} & \cellcolor{orange!90}{6} & \cellcolor{black!75} & \cellcolor{black!75} & \cellcolor{black!75}\\ %\hline
        
        \cellcolor{black!10}{$\bm{D_{1,2/3}}$} & \multicolumn{7}{c}{\cellcolor{black!10}{$[[N, 16, D_{1,2/3}]]$}} \\ %\hline

        \cellcolor{black!10}{$4$} & \cellcolor{orange!10}{$N/6$} & \cellcolor{orange!10}{$\varnothing$} & \cellcolor{orange!90}{4} & \cellcolor{black!75} & \cellcolor{black!75} & \cellcolor{black!75} & \cellcolor{black!75}\\ %\hline

        \cellcolor{black!10}{$12$} & \cellcolor{orange!10}{$N/6$} & \cellcolor{orange!10}{$\varnothing$} & \cellcolor{orange!30}{${N/9}^*$} & \cellcolor{orange!60}{24} & \cellcolor{orange!60}{24} & \cellcolor{orange!75}{20} & \cellcolor{orange!90}{12} \\ %\hline
        
    \end{tabular}
\end{table}

\subsection{Open boundary conditions}\label{subsection:Open boundary conditions}

The code families in \cref{tab:Romanesco codes rectangular} are defined on the torus. We now define them on the cylinder manifold. Each sub-lattice has height $H$ and width $L$ and is periodic at the lateral edges. The number of encoded qubits in the quantum code strongly depends on how we modify the checks at the boundary. Here we considered one simple approach: we truncate the stabilizers at the boundary (see  \cref{app: boundary conditions} for illustrations) and keep those that have weight bigger or equal to 4. Both stabilizer shapes in \cref{tab:Romanesco codes cylinder} yield weight-8 stabilizers in the bulk and weight-6 stabilizers on the top and bottom boundaries. This approach does not always guarantee that the all the stabilizers commute, however they do for those two stabilizer shapes. We still have two stabilizers per plaquette, both in the bulk and everywhere on the boundary. The code parameters are shown in \cref{tab:Romanesco codes cylinder} and are similar to those in \cref{tab:Romanesco codes rectangular}, except that the number of encoded qubits is reduced by half. This is expected since the Romanesco codes have translation invariant stabilizers which are local within some radius, and therefore are akin to other topological codes such as the toric code.

We also consider the two-dimensional plane with open boundaries, in which case we truncate the stabilizers on the left and right boundaries as well. We find that one of the two stabilizer shapes from the cylinder yields commuting stabilizers with this approach, as long as we add two additional data qubits, one gray and one black, at the bottom right and top left corners of the lattice (see  \cref{app: boundary conditions}) so that the weight of the truncated stabilizers at those corners is 6 instead of 5. The stabilizers in the bulk still have weight 8 and all stabilizers on the boundary have weight 6 except four stabilizers (two at the bottom left corner and top right corner) which have weight 4. We list instances of Romanesco codes on a two-dimensional plane in \cref{tab:Romanesco codes 2d plane} with equal height and width, since the quantum distance is given by the minimum dimension. Like on the torus and cylinder, the all-Z logical representatives are supported by the cellular automaton code on each sub-lattice. We find that the number of logical qubits in \cref{tab:Romanesco codes 2d plane} is reduced to four instead of six like on the cylinder manifold in \cref{tab:Romanesco codes cylinder}. It is possible that a different approach to opening the boundary can improve the code parameters. The approach we take here preserves the self-duality of the code; however, we could also consider breaking the self-dual property. For example, we could adapt the code patch using Pauli boundaries instead~\cite{Kesselring2018,Thomsen2024,steffan2025,liang2025planar} or with multiple cellular automaton rules~\cite{Ruiz2025}. We leave a more thorough search of Romanesco codes with open boundary conditions for future work.

For each type of boundary conditions, we observe that the infinite bias distance $d_c$ scales linearly with the number of data qubits $n$ (see \cref{app: boundary conditions}). For example, we find that if $d$ is an integer multiple of $12$ then the quantum code has four logical qubits with an infinite bias distance that is bigger than $n/8$. This differs from the rotated square XZZX code, where the infinite bias distance instead scales with $\sqrt{n}$ with open boundary conditions. This shows how these codes, which have limited-range gates, can result in significant reduction of both the number of qubits and the logical error rate for near-term biased-noise architectures. Additionally, we provide an analysis of the mixed logical representatives in \cref{app: boundary conditions} with open boundary conditions similar to what was done on the torus in \cref{tab:mixed logicals families}.

\begin{table}[t!]
    \centering
    \caption{Families of Romanesco codes on a bipartite honeycomb lattice, where each sub-lattice lies on a cylinder with height $H-1$ and width $L$ so that there are $N^\prime=2 (H-1)L$ data qubits in the quantum codes. We also define the quantities $N = 2 HL$ and $D_{\mu,\nu} = \min(\mu H, \nu L)$ as in \cref{tab:Romanesco codes rectangular}. We get the same parameters as for the codes on the torus except for the number of encoded qubits, which is reduced by half, reducing $v_\infty = k d_c/n$ by half as well.}
    \label{tab:Romanesco codes cylinder}
    \begin{tabular}{c|c|c|c|c|c}
        %\hline
        % \rowcolor{lightgray}
        \cellcolor{black!10}{$\bm{[[n,k,d]]}_\clubsuit^{\cylinder}$} & \cellcolor{black!10}{$\bm{d_c}$} & \cellcolor{black!10}{$\bm{v_\infty}$} & \cellcolor{black!10}{\textbf{Constraints}} & \cellcolor{black!10}{$\bm{R_1}$} & \cellcolor{black!10}{$\bm{R_2}$}\\ \hline
    
        \cellcolor{orange!0}{$[[N^\prime,6,D_{1,1}]]$} & 
        \cellcolor{orange!0}{$\dfrac{3N}{16}$} & $1.13$ & 
        \cellcolor{orange!0}{$\substack{H\%4=0, \ H \geq 4 \\ L\%4=0, \ L \geq 4 \\ H \% 8 \neq 0 \text{ or } L \% 8 \neq 0}$} & 
        \cellcolor{orange!0}{\begin{tikzpicture}[baseline=0.5em]
        \draw[orange!0, very thick] (0,0.6) rectangle (0,0.7);
        \draw[black, very thick, fill=lightgray] (0.2,0) rectangle (0.4,0.2);
        \draw[black, very thick, fill=lightgray] (0.2,0.2) rectangle (0.4,0.4);
        \draw[black, very thick, fill=lightgray] (0.4,0.2) rectangle (0.6,0.4);
        \draw[black, very thick, fill=lightgray] (0,0.4) rectangle (0.2,0.6);
        \draw[orange!0, very thick] (0,-0.1) rectangle (0,0.0);
        \end{tikzpicture}} &
        \cellcolor{orange!0}{
        \begin{tikzpicture}[baseline=0.5em]
        \draw[orange!0, very thick] (0,0.6) rectangle (0,0.7);
        \draw[black, very thick, fill=lightgray] (0.4,0) rectangle (0.6,0.2);
        \draw[black, very thick, fill=lightgray] (0,0.2) rectangle (0.2,0.4);
        \draw[black, very thick, fill=lightgray] (0.2,0.2) rectangle (0.4,0.4);
        \draw[black, very thick, fill=lightgray] (0.2,0.4) rectangle (0.4,0.6);
        \draw[orange!0, very thick] (0,-0.1) rectangle (0,0.0);
        \end{tikzpicture}}\\ \hline

        \cellcolor{orange!0}{$[[N^\prime, 8, D_{1,2/3}]]$} & 
        \cellcolor{orange!0}{$\dfrac{N}{6}$} & $1.33$ & 
        \cellcolor{orange!0}{$\substack{H\% 12 =0, \ H\geq 12 \\ L\%6=0, \ L\geq 6 \\ L\% 12 \neq 0}$} & 
        \cellcolor{orange!0}{\begin{tikzpicture}[baseline=0.5em]
        \draw[orange!0, very thick] (0,0.6) rectangle (0,0.7);
        \draw[black, very thick, fill=lightgray] (0,0) rectangle (0.2,0.2);
        \draw[black, very thick, fill=lightgray] (0.2,0.2) rectangle (0.4,0.4);
        \draw[black, very thick, fill=lightgray] (0.4,0.2) rectangle (0.6,0.4);
        \draw[black, very thick, fill=lightgray] (0,0.4) rectangle (0.2,0.6);
        \draw[orange!0, very thick] (0,-0.1) rectangle (0,0.0);
        \end{tikzpicture}} &
        \cellcolor{orange!0}{\begin{tikzpicture}[baseline=0.5em]
        \draw[orange!0, very thick] (0,0.6) rectangle (0,0.7);
        \draw[black, very thick, fill=lightgray] (0.4,0) rectangle (0.6,0.2);
        \draw[black, very thick, fill=lightgray] (0,0.2) rectangle (0.2,0.4);
        \draw[black, very thick, fill=lightgray] (0.2,0.2) rectangle (0.4,0.4);
        \draw[black, very thick, fill=lightgray] (0.4,0.4) rectangle (0.6,0.6);
        \draw[orange!0, very thick] (0,-0.1) rectangle (0,0.0);
        \end{tikzpicture}}\\ \hline
        
    \end{tabular}
\end{table}

\begin{table}[t!]
    \centering
    \caption{Examples of distance-$d$ Romanesco codes on a bipartite honeycomb lattice with open boundary conditions, where each sub-lattice lies on a two-dimensional plane with height $H=d-1$ and width $L=d-1$, so that each code has $n = 2HL + 2$ data qubits. We show examples up to 1060 data qubits ($4 \leq d \leq 24$) with $v_\infty = kd_c/n \geq 0.4$ and $d\geq 8$. (See \cref{app: boundary conditions} for more examples.)}
    \label{tab:Romanesco codes 2d plane}
    \begin{tabular}{c|c}
        %\hline
        % \rowcolor{lightgray}
        \cellcolor{black!10}{$\bm{[[n,k,d]]}_\clubsuit^{\square}$} & 
        \cellcolor{black!10}{\textbf{Examples}} 
        % & 
        % \cellcolor{black!10}{$\bm{R_1}$} & 
        % \cellcolor{black!10}{$\bm{R_2}$}
        \\ \hline

        \begin{tabular}{c}
            \\[-.5em]
             \hspace{.5em} $[[2(d-1)^2+2, 4, d]]$ \hspace{.5em}\\
             \\[-.5em]
              \begin{tabular}{c|c}
                 \cellcolor{black!10}{\hspace{1em}$\bm{R_1}$\hspace{1em}} & 
        \cellcolor{black!10}{\hspace{1em}$\bm{R_2}$\hspace{1em}} \\
                \hline
                \cellcolor{orange!0}{\begin{tikzpicture}[baseline=0.5em]
                \draw[orange!0, very thick] (0,0.6) rectangle (0,0.7);
                \draw[black, very thick, fill=lightgray] (0.2,0) rectangle (0.4,0.2);
                \draw[black, very thick, fill=lightgray] (0.2,0.2) rectangle (0.4,0.4);
                \draw[black, very thick, fill=lightgray] (0.4,0.2) rectangle (0.6,0.4);
                \draw[black, very thick, fill=lightgray] (0,0.4) rectangle (0.2,0.6);
                \draw[orange!0, very thick] (0,-0.1) rectangle (0,0.0);
                \end{tikzpicture}} &
                \cellcolor{orange!0}{
                \begin{tikzpicture}[baseline=0.5em]
                \draw[orange!0, very thick] (0,0.6) rectangle (0,0.7);
                \draw[black, very thick, fill=lightgray] (0.4,0) rectangle (0.6,0.2);
                \draw[black, very thick, fill=lightgray] (0,0.2) rectangle (0.2,0.4);
                \draw[black, very thick, fill=lightgray] (0.2,0.2) rectangle (0.4,0.4);
                \draw[black, very thick, fill=lightgray] (0.2,0.4) rectangle (0.4,0.6);
                \draw[orange!0, very thick] (0,-0.1) rectangle (0,0.0);
                \end{tikzpicture}}\\
              \end{tabular} 
              \\
              \\[-.5em]
        \end{tabular} & 
        
        \begin{tabular}{c}
        \\[-.7em]
        \begin{tabular}{c | c | c}
         \cellcolor{black!10}{\hspace{.3em} $\bm{d}$ \hspace{.3em}} & \cellcolor{black!10}{\hspace{.3em} $\bm{d_c}$ \hspace{.3em}} & \cellcolor{black!10}{\hspace{.3em} $\bm{v_\infty}$ \hspace{.3em}} \\
         \hline
         % 4 & 4 & 0.80 \\
         \cellcolor{orange!0}{8} & \cellcolor{orange!0}{16} & \cellcolor{orange!0}{0.64} \\
         \cellcolor{orange!0}{12} & \cellcolor{orange!0}{40} & \cellcolor{orange!0}{0.66} \\
         % 16 & 36 & 0.32 \\
         % 20 & 40 & 0.22 \\
         \cellcolor{orange!0}{24} & \cellcolor{orange!0}{160} & \cellcolor{orange!0}{0.60}\\
        \end{tabular}
        \\
        \\[-.7em]
        \end{tabular}
        % & 
        % \cellcolor{orange!0}{\begin{tikzpicture}[baseline=0.5em]
        % \draw[orange!0, very thick] (0,0.6) rectangle (0,0.7);
        % \draw[black, very thick, fill=lightgray] (0.2,0) rectangle (0.4,0.2);
        % \draw[black, very thick, fill=lightgray] (0.2,0.2) rectangle (0.4,0.4);
        % \draw[black, very thick, fill=lightgray] (0.4,0.2) rectangle (0.6,0.4);
        % \draw[black, very thick, fill=lightgray] (0,0.4) rectangle (0.2,0.6);
        % \draw[orange!0, very thick] (0,-0.1) rectangle (0,0.0);
        % \end{tikzpicture}} &
        % \cellcolor{orange!0}{
        % \begin{tikzpicture}[baseline=0.5em]
        % \draw[orange!0, very thick] (0,0.6) rectangle (0,0.7);
        % \draw[black, very thick, fill=lightgray] (0.4,0) rectangle (0.6,0.2);
        % \draw[black, very thick, fill=lightgray] (0,0.2) rectangle (0.2,0.4);
        % \draw[black, very thick, fill=lightgray] (0.2,0.2) rectangle (0.4,0.4);
        % \draw[black, very thick, fill=lightgray] (0.2,0.4) rectangle (0.4,0.6);
        % \draw[orange!0, very thick] (0,-0.1) rectangle (0,0.0);
        % \end{tikzpicture}}
        \\ \hline

        \begin{tabular}{c}
            \\[-.5em]
             $[[2(d-1)^2+2, 2, d]]$ \\
             \\[-.5em]
              \begin{tabular}{c|c}
                 \cellcolor{black!10}{\hspace{1em}$\bm{R_1}$\hspace{1em}} & 
        \cellcolor{black!10}{\hspace{1em}$\bm{R_2}$\hspace{1em}} \\
                \hline
                \cellcolor{orange!0}{\begin{tikzpicture}[baseline=0.5em]
                \draw[orange!0, very thick] (0,0.6) rectangle (0,0.7);
                \draw[black, very thick, fill=lightgray] (0.2,0) rectangle (0.4,0.2);
                \draw[black, very thick, fill=lightgray] (0.2,0.2) rectangle (0.4,0.4);
                \draw[black, very thick, fill=lightgray] (0.4,0.2) rectangle (0.6,0.4);
                \draw[black, very thick, fill=lightgray] (0,0.4) rectangle (0.2,0.6);
                \draw[orange!0, very thick] (0,-0.1) rectangle (0,0.0);
                \end{tikzpicture}} &
                \cellcolor{orange!0}{
                \begin{tikzpicture}[baseline=0.5em]
                \draw[orange!0, very thick] (0,0.6) rectangle (0,0.7);
                \draw[black, very thick, fill=lightgray] (0.4,0) rectangle (0.6,0.2);
                \draw[black, very thick, fill=lightgray] (0,0.2) rectangle (0.2,0.4);
                \draw[black, very thick, fill=lightgray] (0.2,0.2) rectangle (0.4,0.4);
                \draw[black, very thick, fill=lightgray] (0.2,0.4) rectangle (0.4,0.6);
                \draw[orange!0, very thick] (0,-0.1) rectangle (0,0.0);
                \end{tikzpicture}}\\
              \end{tabular}
              \\
              \\[-.5em]
        \end{tabular} & 
        \cellcolor{orange!0}{\hspace{.5em}
        \begin{tabular}{c}
        \\[-.7em]
        \begin{tabular}{c | c | c}
             \cellcolor{black!10}{\hspace{.3em} $\bm{d}$ \hspace{.3em}} & \cellcolor{black!10}{\hspace{.3em} $\bm{d_c}$ \hspace{.3em}} & \cellcolor{black!10}{\hspace{.3em} $\bm{v_\infty}$ \hspace{.3em}} \\
             \hline
             % \cellcolor{orange!0}{5} & \cellcolor{orange!0}{8} & \cellcolor{orange!0}{0.47}\\
             % 6 & 10 & 0.38\\
             % \cellcolor{orange!0}{7} & \cellcolor{orange!0}{19} & \cellcolor{orange!0}{0.51}\\
             \cellcolor{orange!0}{9} & \cellcolor{orange!0}{32} & \cellcolor{orange!0}{0.49}\\
             % 10 & 30 & 0.37\\
             % 11 & 36 & 0.36\\
             13 & 60 & 0.41\\
             14 & 72 & 0.42\\
             % 15 & 67 & 0.34\\
             17 & 104 & 0.40\\
             % 18 & 74 & 0.26\\
             % 19 & 76 & 0.23\\
             % 21 & 80 & 0.20\\
             % 22 & 82 & 0.19
        \end{tabular}
        \\
        \\[-.7em]
        \end{tabular}
        \hspace{.5em}
        } 
        %& 
        % \cellcolor{orange!0}{\begin{tikzpicture}[baseline=0.5em]
        % \draw[orange!0, very thick] (0,0.6) rectangle (0,0.7);
        % \draw[black, very thick, fill=lightgray] (0.2,0) rectangle (0.4,0.2);
        % \draw[black, very thick, fill=lightgray] (0.2,0.2) rectangle (0.4,0.4);
        % \draw[black, very thick, fill=lightgray] (0.4,0.2) rectangle (0.6,0.4);
        % \draw[black, very thick, fill=lightgray] (0,0.4) rectangle (0.2,0.6);
        % \draw[orange!0, very thick] (0,-0.1) rectangle (0,0.0);
        % \end{tikzpicture}} &
        % \cellcolor{orange!0}{
        % \begin{tikzpicture}[baseline=0.5em]
        % \draw[orange!0, very thick] (0,0.6) rectangle (0,0.7);
        % \draw[black, very thick, fill=lightgray] (0.4,0) rectangle (0.6,0.2);
        % \draw[black, very thick, fill=lightgray] (0,0.2) rectangle (0.2,0.4);
        % \draw[black, very thick, fill=lightgray] (0.2,0.2) rectangle (0.4,0.4);
        % \draw[black, very thick, fill=lightgray] (0.2,0.4) rectangle (0.4,0.6);
        % \draw[orange!0, very thick] (0,-0.1) rectangle (0,0.0);
        % \end{tikzpicture}}
        \\ \hline
        
    \end{tabular}
\end{table}

\section{Numerical results} \label{sec: numerical results}

Here we report the total logical error rate (defined as the average of the X and Z logical error rates) under code-capacity noise for different Romanesco codes. We work with a basis of logical representatives where Z (X) logical operators are composed of all Z (X) Pauli operators (up to stabilizers) in the non-CSS code. We solve the decoding problem with the CSS code $C$ but rotate the noise channels so that we effectively simulate $C_\clubsuit$ (see \cref{app:rotating decoding problem}).

\begin{figure}[t!]
    \centering
    \includegraphics[width=\linewidth]{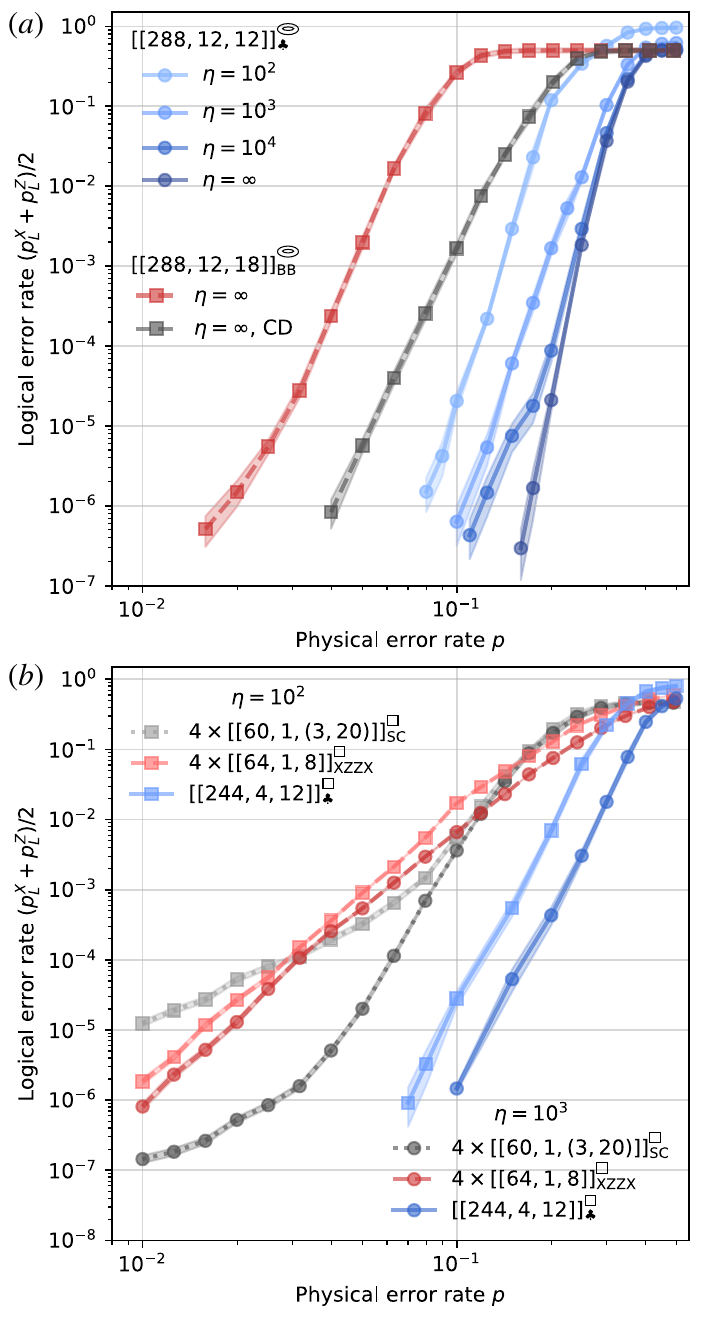}
    \caption{Total logical error rate (average of the X and Z logical error rate) for code-capacity noise as a function of the physical error rate $p=p_Z$ for different values of noise bias $\eta = p_Z/p_X$, for (a) the $[[288, 12, 12]]_\clubsuit^{\protect \torus}$ code with infinite bias distance $d_c=54$ and the bivariate bicycle code $[[288,12,18]]_{\rm BB}^{\protect \torus}$ (where `CD' in the legend stands for Clifford-deformed using \cref{eq:HadamardRotation}~\cite{Roffe2023}) and (b) the $[[244, 4, 12]]_\clubsuit^\square$ and four copies of the rotated square XZZX code $[[64, 1, 8]]_{\rm XZZX}^\square$ and four copies of the rectangular rotated surface code $[[60, 1, (3, 20)]]_{\rm SC}^\square$ (with a Z distance larger than the X distance) so that the number of data qubits is approximately the same for all curves. The shaded regions identify the 95\% confidence intervals for a Poisson distribution obtained with SciPy~\cite{scipy}.}
    \label{fig:memory}
\end{figure}

In \cref{fig:memory} we focus on two examples of Romanesco codes: $[[288, 12, 12]]_{\clubsuit}^{\torus}$ (part of the $[[N,12,D_{1,1}]]$ family on the torus with infinite bias distance $d_c=54$ and dimensions $H=L=12$) and $[[244, 4, 12]]_{\clubsuit}^{\square}$ (on the 2D plane with $d_c=40$ and $H=L=11$). We compare $[[288, 12, 12]]_{\clubsuit}^{\torus}$ with the bivariate bicycle code $[[288, 12, 18]]_{\rm BB}^{\torus}$~\cite{Bravyi2024} and $[[244, 4, 12]]_{\clubsuit}^{\square}$ with four copies of 1) the rotated square XZZX code $[[64, 1, 8]]_{\rm XZZX}^{\square}$ and 2) the thin surface code $ [[60, 1, (3, 20)]]_{\rm SC}^{\square}$, where $d_X = 3$ and $d_Z=20$ such that the code offers greater protection against phase-flip errors than bit-flip errors. This comparison keeps the number of logical and physical qubits approximately equal. Each data point includes at least 10 sampled logical errors.

We used the tesseract decoder~\cite{beni2025tesseract} for $[[288, 12, 12]]_{\clubsuit}^{\torus}$ and $[[244, 4, 12]]_{\clubsuit}^{\square}$ with beam climbing and a maximum beam factor 40. Whenever a sample could not be decoded within one minute, we used a modified BP+OSD decoder instead (see \cref{app: custom decoder}). In the infinite bias limit we use BP+OSD as proposed by Panteleev and Kalachev~\cite{Panteleev2021degeneratequantum} (labeled as BP+OSD-E in \cite{Roffe2020}) on the classical codes supported on the black and gray sub-lattices with 1000 BP iterations and OSD order $k_c=7$ and $2$ for the $[[288, 12, 12]]_{\clubsuit}^{\torus}$ and $[[244, 4, 12]]_{\clubsuit}^{\square}$ codes, respectively\footnote{The maximum OSD order when solving the decoding problem with BP+OSD and a $m\times n$ parity check matrix $H$ is $n - \mathrm{rank}(H)$, which is exactly the number of encoded qubits for a classical code.}. We used the BP+OSD-CS decoder~\cite{Roffe2020,Roffe_LDPC_Python_tools_2022} (product-sum BP with 1000 maximum iterations and OSD order 50) for the bivariate bicycle code $[[288, 12, 18]]_{\rm BB}^{\torus}$ in (a). We separated the X and Z decoding graphs for speed and saw no significant advantage when using the full decoding graph under code-capacity noise. We tested the tesseract decoder for the $[[288, 12, 18]]_{\rm BB}^{\torus}$ bivariate bicycle code in~\cref{app: decoder comparison} and found that it did not improve the logical error rate for infinite bias. We used minimum-weight perfect matching (PyMatching)~\cite{Higgott2025sparseblossom} for the XZZX and thin surface codes in (b).

Focusing on $[[288, 12, 12]]_{\clubsuit}^{\torus}$ in panel (a), the $\eta=\infty$ curve has an effective distance $54$ instead of $12$. However, we observe that the slope for moderate noise bias $\eta = 10^2$ and $10^3$ is lower than $54$. We expect an effective distance of roughly $36$ because the shortest mixed logical representatives have weight $36$ in \cref{tab:mixed logicals families}. For $\eta = 10^4$, we observe a sudden change in the slope where the effective distance should decrease from 54 to 36. The curve could potentially be smoothened out further by using a larger timeout time for the tesseract decoder. Despite the fact that $[[288, 12, 18]]_{\rm BB}^{\torus}$ has a higher code distance than our $[[288, 12, 12]]_{\clubsuit}^{\torus}$ code, our code offers a significantly improved logical performance for the same encoding rate in the moderate to large bias regime. Below some small physical error rate (set by the noise bias and depending on the decoder) $[[288, 12, 18]]_{\rm BB}^{\torus}$ could eventually outperform $[[288, 12, 12]]_{\clubsuit}^{\torus}$ due to a larger code distance. However, Clifford-deforming $[[288, 12, 18]]_{\rm BB}^{\torus}$ using Hadamard rotations like in \cref{eq:HadamardRotation} only yields an infinite bias distance 18. As a result, we find that $[[288, 12, 12]]_{\clubsuit}^{\torus}$ significantly outperforms $[[288, 12, 18]]_{\rm BB}^{\torus}$ for moderate values of bias and physical error rate.

Moving on to $[[244, 4, 12]]_{\clubsuit}^{\square}$ in panel (b), we observe that the logical performance is improved relative to both the XZZX and thin surface codes, even in the moderate bias regime. Here we remark that the minimum weight of a logical representative with a single X operator in our code is $22$ (see \cref{app: boundary conditions} for more details about the low-weight logical operators) so that our $[[244, 4, 12]]_{\clubsuit}^{\square}$ code has an effective distance closer to $22$, eventually reducing to $12$ at low physical error rates $p_Z$. For the $[[60, 1, (3,20)]]_{\rm SC}^\square$, the effective code distance is 20 at large bias and eventually decreases to 3 for small $p_Z$. The $[[64, 1, 8]]_{\rm XZZX}^\square$ code keeps the same effective distance $8$ for all physical error rates. In this plot we report of the logical error rate of k separate patches of the surface code as $1 - (1-p_{L1})^k$ where $p_{L1}$ is the logical error probability of a single patch.

More generally, a Romanesco code $[[n,k,d]]_\clubsuit^\square$ on the 2D plane (see \cref{tab:Romanesco codes 2d plane}) has $n=2(d-1)^2+2$ data qubits and an infinite bias distance $d_c = 2 \alpha d^2$, where $1/24 \leqslant \alpha \leqslant 1/8$ for $k=4$ and $1/8 \leqslant \alpha \leqslant 1/4$ for $k=2$ (see \cref{app: boundary conditions}). The largest square rotated XZZX code with at most $n/k$ data qubits has parameters $[\tilde{d}^2,1,\tilde{d}]]_{\mathrm{XZZX}}$ where $\tilde{d} = \lfloor (d-1)/\sqrt{k/2}\rfloor$. Similarly, the largest rectangular rotated surface codes with at most $n/k$ data qubits have parameters $[[\tilde{d}_X \tilde{d}_Z, 1, (\tilde{d}_X, \tilde{d}_Z)]]_{SC}$ where $\tilde{d}_Z = \lfloor \tilde{d}^2/d_X \rfloor$ is the Z distance and $\tilde{d}_X \leq \tilde{d}_Z$ is the X distance. 

There are two physical error rate regimes of interest with fixed bias. In the limit of small error rate $p_Z$ the quantum code distance $d$ determines the exponent of the logical error rate $p_L \propto p_Z^{\lfloor d /2\rfloor}$. In that regime, the Romanesco code always outperforms the XZZX and rectangular surface codes since $d > \tilde{d} \geq \tilde{d}_X$. Below the bias-dependent transition point where the effective code distance reduces to the size of the lowest-weight logical (which is half X and half Z for the Clifford-deformed codes), the logical error rate is also suppressed with respect to $\eta$ as follows: $p_L \propto \eta^{-\lfloor d/4 \rfloor}  p_Z^{\lfloor d/2\rfloor}$ for the Romanesco codes, $p_L \propto \eta^{-\lfloor \tilde{d}/4 \rfloor}  p_Z^{\lfloor \tilde{d}/2\rfloor}$ for the XZZX code and $p_L\propto \eta^{-\lfloor \tilde{d}_X/2 \rfloor}  p_Z^{\lfloor \tilde{d}_X/2\rfloor}$ for the rectangular surface code. 

In the moderate to large error rate regime the effective code distance $d^\prime$ is $d_c$ for Romanesco codes, $\tilde{d}$ for the square XZZX code and $\tilde{d}_Z$ for the rectangular surface code. The logical error rate $p_L \propto p_Z^{\lfloor d^\prime/2 \rfloor}$ for each code. It is possible to find $\tilde{d}_Z \geq d_c$ if $\alpha \leq (k\tilde{d}_X)^{-1}$; however, we generally find that $d_c$ is comparable or higher than $\tilde{d}_Z = \tilde{d}^2/3$ (where $\tilde{d}_X =3$). Therefore, we find that our codes generally outperform rotated surface codes, both the XZZX and rectangular variants, in all regimes of noise bias and physical error rates.

\section{Conclusion} \label{sec: conclusion}

In this work we introduced a family of self-dual Clifford-deformed bivariate bicycle codes tailored to biased noise, called Romanesco codes, with limited-range gates and weight-8 stabilizers. The stabilizer generators are each half X and half Z, similar to other bias-tailored codes proposed in the literature~\cite{Tuckett2019,Roffe2023}. In the infinite bias limit, the decoding graph reduces to the decoupled graphs of the two input classical cellular automaton codes. These classical codes were recently proposed and studied with cat qubits~\cite{Ruiz2025}, since they offer better encoding rates than the repetition code while having quasi-local checks. Importantly, the distance of these classical codes can scale linearly with the number of data qubits no matter the choice of boundary conditions. As a result, the logical error rate of our codes is suppressed by orders of magnitude under code-capacity noise in the presence of strong noise bias, where the effective code distance is approximately that of the input classical codes. Our codes also yield higher encoding rates than the surface code.

Inspired by the bias-tailored honeycomb color code, we choose to lay out our codes on a bipartite hexagonal lattice, where each sub-lattice supports one of the input classical codes. However, other layouts could be considered. It is possible to reduce the connectivity requirements by using two ancilla qubits per plaquette in the following way: each ancilla qubit is coupled to its own sub-lattice so that each ancilla qubit is connected to $w/2$ data qubits. The two ancilla qubits also need to be linked to measure the full stabilizers. This would result in a connectivity $w/2$ on data qubits and $w/2+1$ on ancilla qubits. This scheme could be particularly relevant for the transmon-cat architecture~\cite{hann2024}. We leave the optimization of the syndrome extraction circuit and the simulation of the memory experiment with circuit-level noise for future work. However, we believe that these codes could be well suited for near-term realization of qLDPC codes with biased-noise qubits if next-nearest neighbor two-qubit gates~\cite{chandra2024} were realizable on each sub-lattice.

Finally, we remark that decoders that make use of cellular automaton rules could be used to improve the logical performance of our codes ~\cite{Miguel2023,Vasmer2021}.

\section{Acknowledgments}

We thank Fernando Pastawski, Przemyslaw Bienias, Sophia Lin, Esha Swaroop, Arne Grimsmo, Yunong Shi, Will Morong and Fernando Brand\~{a}o for fruitful discussions. We also thank Oskar Painter, Simone Severini, Bill Vass, James Hamilton, Nafea Bshara, Peter DeSantis, and Andy Jassy at Amazon, for their involvement and support of the research activities at the AWS Center for Quantum Computing.

\appendix

\section{From the color code to decoupled cellular automaton codes}\label{app:color code to fractal codes}

\begin{figure}[h!]
    \centering
    \includegraphics[width=\linewidth]{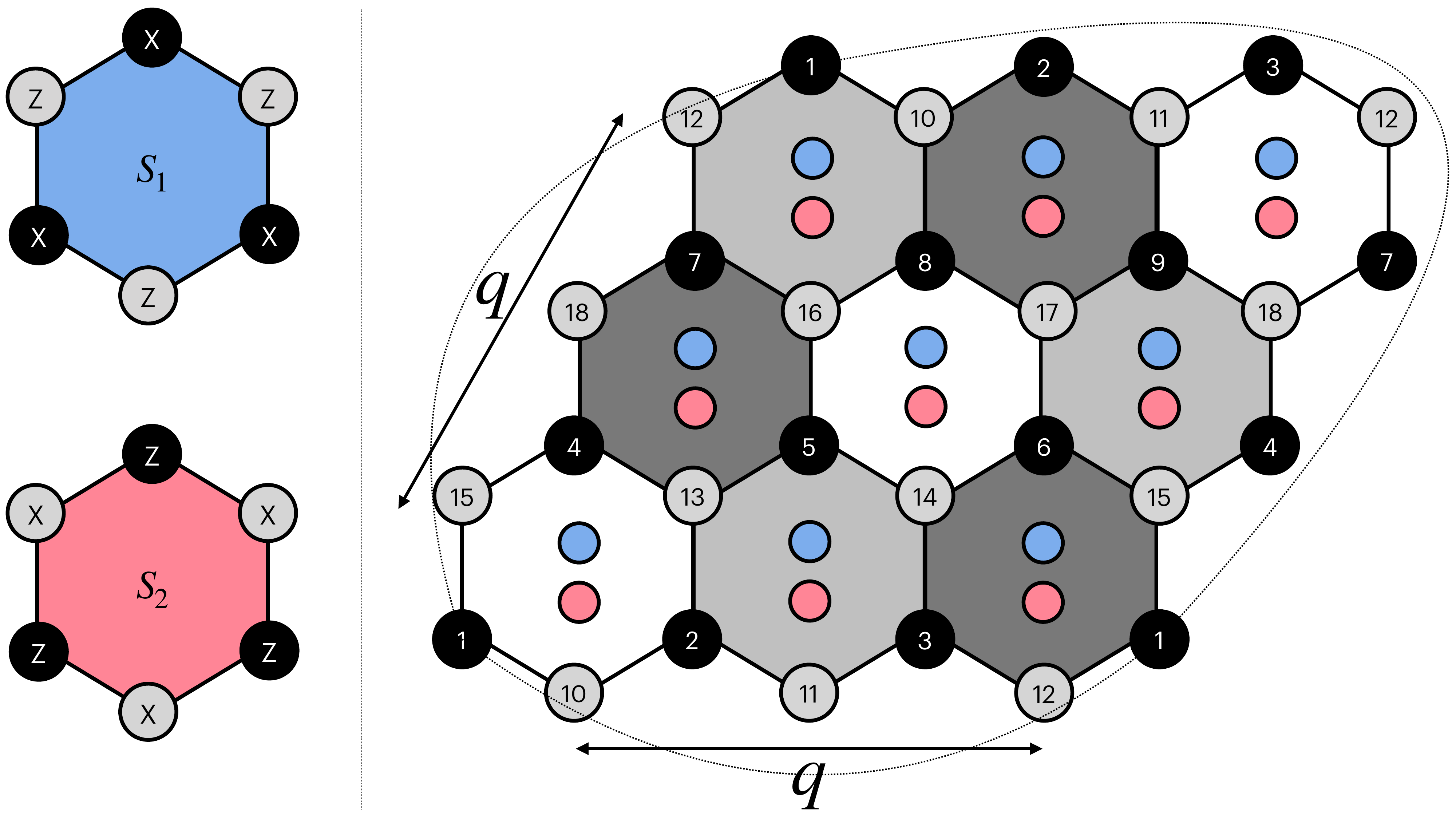}
    \caption{The XZ$^3$ code.}
    \label{fig:xyz code}
\end{figure}

In \cref{fig:xyz code} we show the $[[18q^2, 4, 4q]]$ color code~\cite{deCarvalho2021} on a bipartite honeycomb lattice. The data qubits are identified by black and gray circles. The lattice has periodic boundary conditions so that qubits labeled with the same numbers are the same qubits. Here $q$ is the number of times each color (white, light and dark gray) repeats along each boundary. The code has $9q^2$ plaquettes and there are two weight-6 stabilizers, $S_1$ and $S_2$, per plaquette. $S_1$ ($S_2$) is composed of X (Z)-type parity checks on the three black qubits and the three gray qubits on the plaquette. In this work we propose to apply a Hadamard rotation (similarly to what was done in the XYZ color code~\cite{Miguel2023}) on the gray qubits such that both $S_1$ and $S_2$ now have alternating X- and Z-type parity checks as illustrated in \cref{fig:xyz code}: the X-type checks are on the black (gray) qubits for $S_1$ ($S_2$).

\begin{figure}[h!]
    \centering
    \includegraphics[width=\linewidth]{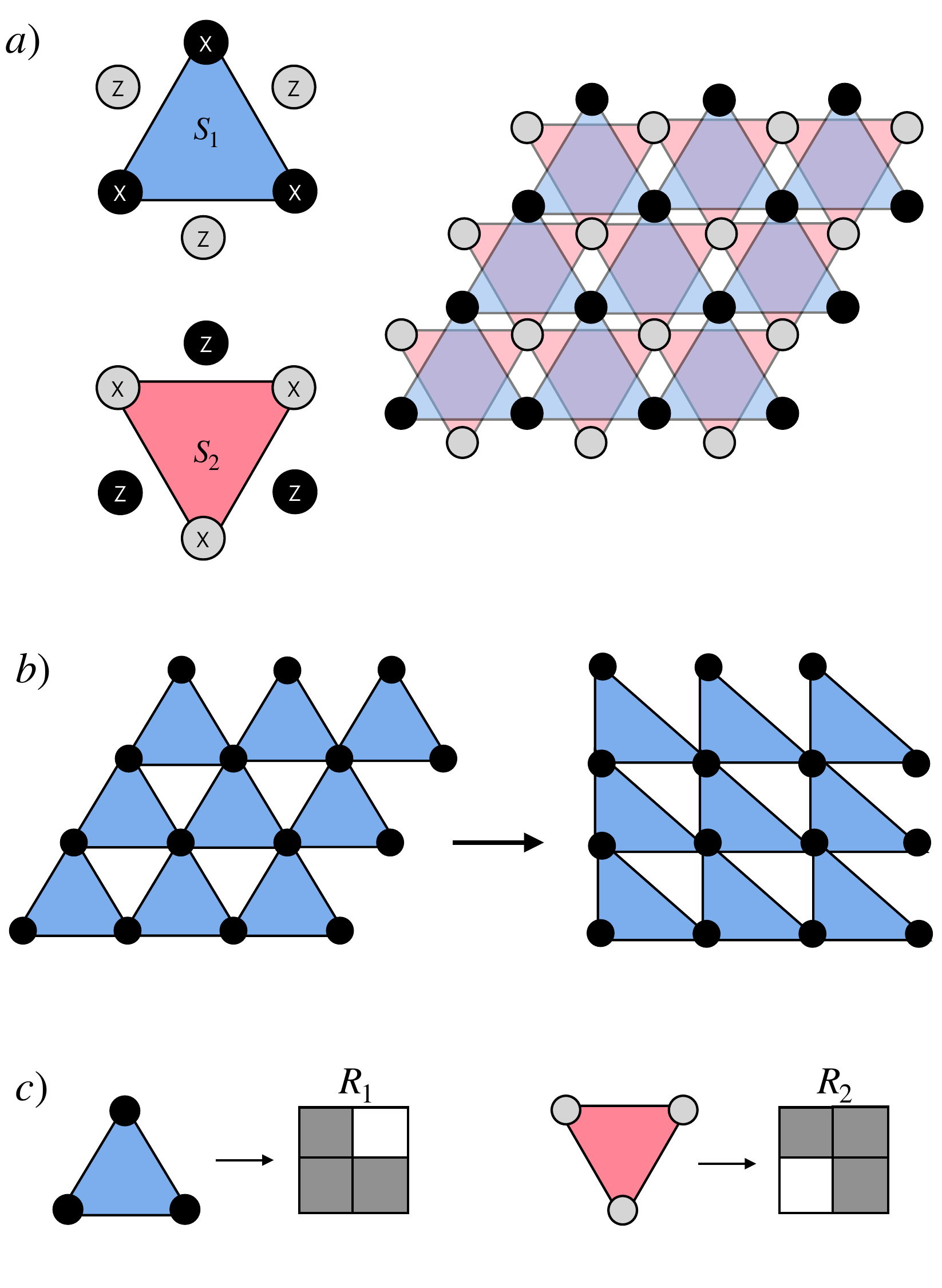}
    \caption{(a) The XZ$^3$ code in the infinite bias limit, i.e $p_Z \to \infty$ and $p_X, p_Y \to 0$. (b) Going from a hexagonal lattice to a rectangular lattice. (c) Cellular automaton rules for the XZ$^3$ code.}
    \label{fig:xyz ca rules}
\end{figure}

In the limit of large noise bias where phase-flips dominate over bit-flips, i.e. $p_Z\gg p_X$, we can approximately ignore the Z-checks in the decoding graph as explained in the main text. Removing the Z-checks from the stabilizers $S_1$ and $S_2$ result in two decoupled decoding graphs with weight-3 X-type stabilizers, one acting on the black qubits and the other on the gray qubits, as sketched in \cref{fig:xyz ca rules} a). Each decoupled decoding graph is a decoding graph that can be obtained by using a classical cellular automaton code with weight-3 stabilizers. This was first noted in \cite{Miguel2023} where it was shown that in the infinite bias limit we can think of the XYZ color code as two copies of the cellular automaton code with rule 102~\cite{Wolfram1983}. It is easier to recognize this rule by modifying the grid as shown in panel (b) such that the data qubits are now aligned on a rectangular grid. Throughout this work, we identify the cellular automaton rules with binary matrices $R_1$ or $R_2$ for the black or gray sub-lattice, respectively. These binary matrices determine which qubits are part of the stabilizer when looking at the rectangular grid in panel (b) on the right. For example, in panel (c) on the left we show that the blue weight-3 stabilizer we obtained in the color code when looking only at black qubits is rule 102, here labeled with $R_1$. The rule on the gray qubits is always a rotated version of the rule on the black qubits, as shown in panel (c).

\section{Numerical search of Romanesco codes on the torus}\label{app:Numerical search}

In this appendix, we explain how we ultimately found the code families in \cref{tab:Romanesco codes rectangular}. The first step is to identify promising classical codes for the black and gray sub-lattices. The second step is to combine pairs of classical codes as in \cref{fig:construction} to get quantum Romanesco codes that satisfy the required commutation relations. The third step is to identify the best pairs of classical codes (or stabilizer shapes) based on the code parameters of the quantum codes. The fourth and final step is to define code families whose members have similar code parameters.

\subsection{Searching for classical codes}

We generate all unique cellular automaton rules $R$ of size $m\times m$ with weight $w$ in both rows and columns (taking the periodic boundary conditions into account). We also make sure that the bottom row and left column have at least a non-zero entry. For each cellular automaton rule, we generate the parity check matrix with periodic boundary conditions and compute the code parameters. We only keep codes that are not made of decoupled classical codes by checking that the code stabilizers form a single connected component when defining a graph whose edges are the gates in the stabilizers.

Here we consider periodic square lattices for the cellular automaton codes with $9 q^2$ qubits (such that the quantum codes have $18 q^2$ qubits) with $q=1,2,3,4,5$. The square lattices will be transformed into triangular lattices (see \cref{fig:xyz ca rules}). We remark that we chose to only consider honeycomb lattices with $9 q^2$ plaquettes with integer $q$ but our code construction works for honeycomb lattices of arbitrary sizes. We use this geometric constraint to find the most promising stabilizer shapes. We also considered rectangular lattices but did not find better encoding rates with periodic conditions. For each pair of these shapes we will then identify code families with promising code parameters for arbitrary lattice sizes.

Moreover, we only consider $w = 3, 4$ such that we have weight-$w$ stabilizers in the classical code and therefore weight-$2w$ stabilizers in the quantum code, and cellular automaton rule size $m=2,3,4$. For example, the XZ$^3$ code is made of $w=3$ and $m=2$ cellular automaton codes. We numerically found that $m=5$ with $w\leq 4$ did not give better code parameters for the classical codes with periodic boundary conditions. We only keep the classical codes that have at least 4 encoded qubits and distances bigger than 3.

The number of encoded bits in the classical code is $k=n - \mathrm{rank}(H)$ in mod 2 space where $H$ is the parity check matrix of the code. The code distance is found using the same algorithm as for the quantum codes (i.e. \cref{algo:code distance} which uses the decoding problem to find the low-weight logical operators) but we pass a $0 \times n$ matrix for $H_Z$ whose kernel is just the $n\times n$ identity matrix. When using \cref{algo:code distance} we find an upper bound on the code distance which can be made tight by sampling enough error vectors. We check the convergence by generating 100 new low-weight logical operators. We used both the BP+OSD-CS and linear integer program decoder for codes below 500 qubits.

\begin{algorithm}[h!]
    \caption{
    We solve a modified decoding problem to obtain the code distance for any CSS code with parity check matrices $H_X$ and $H_Z$. The roles of $H_X$ and $H_Z$ can be swapped. \\
    Here, $\mathrm{rk}_{\mathrm{mod}2}$ and $\mathrm{ker}_{\mathrm{mod}2}$ refers to the rank and kernel mod 2, respectively, $\mathrm{vstack}(A, B)$ vertically stacks $A$ on top of $B$. $\mathrm{nrow}(A)$ and $\mathrm{ncol}(A)$ is the number of rows and columns of $A$, respectively. \\
    Here the decoder can be either BP+OSD-CS~\cite{Roffe2020} at maximum OSD order (i.e. number of qubits minus rank mod 2 of the parity check matrix) to maximize accuracy, or a linear integer program. In the case of BP+OSD-CS, we use 1 BP iteration with the product-sum algorithm to sort the columns of the parity check matrix for OSD, and use a depolarizing rate~$1e-4$ without any noise bias for every qubit.
    }\label{algo:code distance}
    \begin{algorithmic}
        \Procedure{Code distance}{$H_X$, $H_Z$}
        \State \textbf{input: } X parity check matrix $H_X$, Z parity check matrix $H_Z$
        \State \textbf{output: } The distance of the CSS code
        \State $[L_Z] \gets$ list of Z logicals
        \State // \textit{Get test X logicals}
        \State $[L_X] \gets$ list of X test logicals
        \For{$L_X \in \mathrm{ker}_{\mathrm{mod}2}(H_Z)$}
        \State $\tilde{H}_X = \mathrm{vstack}(H_X, L_X)$
        \State // \textit{check it is not in the rowspace of $H_X$}
        \If{$\mathrm{rk}_{\mathrm{mod}2}(\tilde{H}_X)\neq \mathrm{rk}_{\mathrm{mod}2}(H_X)$}
        \State add $L_X$ to $[L_X]$
        \EndIf
        \EndFor
        \State // \textit{Find low-weight Z logicals}
        \State // $[\tilde{L}_X] \gets$ \textit{random combinations of $[L_X]$}
        \For{$\tilde{L}_X \in [\tilde{L}_X]$}
        \State // \textit{set up the decoder with the same depolarizing} 
        \State // \textit{rate $p$ on all data qubits}
        \State $D_X = \mathrm{BPOSD}(\mathrm{vstack}(H_X,\tilde{L}_X), p)$
        \State $s_X = \mathrm{vstack}(\vec{0}_{\mathrm{nrow}(H_X)}, 1)$ // syndrome
        \State $L_Z = D_X(s_X)$ // \textit{decode to get Z logical}
        \State add $L_Z$ to $[L_Z]$
        \EndFor
        \Return min $[L_Z]$
        \EndProcedure
    \end{algorithmic}
\end{algorithm}

\subsection{Searching for quantum codes}

We do not allow for $d=2$ codes: these cases directly imply that there exists a weight-2 mixed logical operator with one Pauli Z and one Pauli X that connects a black qubit to a gray qubit in $C_\clubsuit$ since the classical code in each sub-lattice strictly have higher weight logical operators. These errors have linear sensitivity in the noise bias, meaning that their performance reduces to that of two decoupled classical LDPC codes. Therefore, in order to speed up the search, we first determine if $R_1$ and $R_2$ result in a valid quantum code with distance greater than 2. We check that the pair of rules does not result in weight-2 logical operators which would linearly sensitive in the noise bias as noted above. We do so by looking that there isn't a weight-2 error that is non-detectable using the parity check matrix of the quantum, obtained by stacking the parity check matrices of both classical codes. We choose a gray qubit to have a bit-flip error and for each black qubit at a time, we add a phase-flip error. We check that all syndromes are non-trivial and distinct from having just the bit-flip or phase-flip error. If true, then we check that stabilizers commute. If the commutation relations are satisfied then we keep the quantum code and compute its code parameters: $k=n - \mathrm{rank}(H_X) - \mathrm{rank}(H_Z)$ in mod 2 space and the code distance is found using \cref{algo:code distance}.

\Cref{algo:periodic code} shows how to build the parity check matrix of a periodic Romanesco code.

\begin{algorithm}[h!]
    \caption{
    Pseudo-code to build a Romanesco code on a bipartite honeycomb lattice with periodic boundary conditions with $2HL$ qubits: each sub-lattice has height $H$ and width $L$. The classical codes on the two sub-lattices are specified by the cellular automaton rules $R_1$ and $R_2$.}\label{algo:periodic code}
    \begin{algorithmic}
        \Procedure{Romanesco Code}{$H$, $L$, $R_1$, $R_2$}
        \State \textbf{input: } Height $H$, width $L$, $m\times m$ cellular automaton rule on black (gray) qubits $R_1$ ($R_2$)
        \State \textbf{output: } Parity check matrix
        \State $h = \boldsymbol{0}_{HL \times 2HL}$ // \textit{X or Z parity check matrix}
        \State $r = 0$
        \For{$0 \leq i < L$} // \textit{for ith column}
        \For{$0 \leq j < H$} // \textit{for jth row}
        \State $[q] = [\phantom{.}]$ // \textit{list of data qubits in stabilizer}
        \For{$0 \leq k < m$} // \textit{kth column of $R_{1(2)}$}
        \For{$0 \leq l < m$} // \textit{lth row of $R_{1(2)}$}
        \State $i^\prime, \ j^\prime =  i + k, \ j + l$
        \State // 
        \If{$[R_1]_{m-1-k,l} = 1$}
        \State $x = \mod(2 i^\prime, \ 2L)$
        \State $y = \mod(2 j^\prime + 1, \ 2H)$
        \State // \textit{the lattice is slanted}
        \State append $(\lfloor y/2 \rfloor + x, \ y)$ to $q$
        \EndIf
        \If{$[R_2]_{m-1-k,l} = 1$}
        \State $x = \mod(2 i^\prime - 1, \ 2L)$
        \State $y = \mod(2 j^\prime, \ 2H)$
        \State append $(\lfloor y/2 \rfloor + x, \ y)$ to $q$
        \EndIf
        \EndFor
        \EndFor
        \For{$q \in [q]$}
        \State get index $s$ for qubit $q$
        \State $h[r, s] = 1$
        \EndFor
        \State $r = r + 1$
        \EndFor
        \EndFor
        \Return $H = 
        \begin{pmatrix}
        h & 0 \\ 0 & h    
        \end{pmatrix}$
        \EndProcedure
    \end{algorithmic}
\end{algorithm}

\subsection{Selecting the best codes}

Next we need to define what are good code parameters for this code construction. Our approach is as follows. For each unique $k$ value we obtained for quantum codes, we keep the quantum codes with the best classical code distance (or infinite bias distance) $d_{\infty}$. Among these codes, we then keep those with the minimum stabilizer weight $w$. We then keep those with the small cell size $m$. For the same lattice size, we then remove the codes that have smaller $k$ and $d_\infty$ than other codes in the group, and for equal $k$ and $d_\infty$ we keep those with the highest distance $d$ (at zero bias). The best codes we found are listed in \cref{tab:Romanesco codes}.

\begin{table}[h!]
    \centering
    \caption{Best Romanesco codes up to 450 data qubits defined on a bipartite honeycomb lattice where each sub-lattice has height $H = 3q$ and width $L = 3q$ with integer $q$ for a total of $18 q^2$ data qubits in the quantum code. For each lattice size we highlighted the code with the highest ratio $v_\infty = kd_c/n$: the code has $v_\infty$ less data qubits than $k$ repetition codes of distance $d_c$. Here the two classical codes composing each quantum code are defined on a torus and so is the resulting quantum code.}
    \label{tab:Romanesco codes}
    \begin{tabular}{c|c|c|c|c|c}
        %\hline
        % \rowcolor{lightgray}
        \cellcolor{black!10}{$\bm{[[n,k,d]]}$} & \cellcolor{black!10}{$\bm{[n_c, k_c, d_c]}$} & \cellcolor{black!10}{$\bm{k/n}$} & \cellcolor{black!10}{$\bm{kd_c/n}$} & \cellcolor{black!10}{${\bm{R_1}}_{\torus}$} & \cellcolor{black!10}{${\bm{R_2}}_{\torus}$}\\ \hline
        
        \cellcolor{orange!10}{$[[72,12,4]]$} & 
        \cellcolor{orange!10}{$[36, 10, 12]$} & 
        \cellcolor{orange!10}{$0.17$} & 
        \cellcolor{orange!10}{$\bm{2.00}$} & 
        \cellcolor{orange!10}{\begin{tikzpicture}[baseline=0.5em]
        \draw[orange!10, very thick] (0,0.6) rectangle (0,0.7);
        \draw[black, very thick, fill=lightgray] (0,0) rectangle (0.2,0.2);
        \draw[black, very thick, fill=lightgray] (0.2,0.2) rectangle (0.4,0.4);
        \draw[black, very thick, fill=lightgray] (0.4,0.2) rectangle (0.6,0.4);
        \draw[black, very thick, fill=lightgray] (0,0.4) rectangle (0.2,0.6);
        \draw[orange!10, very thick] (0,-0.1) rectangle (0,0.0);
        \end{tikzpicture}} &
        \cellcolor{orange!10}{\begin{tikzpicture}[baseline=0.5em]
        \draw[orange!10, very thick] (0,0.6) rectangle (0,0.7);
        \draw[black, very thick, fill=lightgray] (0.4,0) rectangle (0.6,0.2);
        \draw[black, very thick, fill=lightgray] (0,0.2) rectangle (0.2,0.4);
        \draw[black, very thick, fill=lightgray] (0.2,0.2) rectangle (0.4,0.4);
        \draw[black, very thick, fill=lightgray] (0.4,0.4) rectangle (0.6,0.6);
        \draw[orange!10, very thick] (0,-0.1) rectangle (0,0.0);
        \end{tikzpicture}}\\ \hline
        
        $[[72,16,4]]$ & 
        $[36, 10, 6]$ & 
        $0.22$ & 
        $1.33$ & 
        \begin{tikzpicture}[baseline=0.5em]
        \draw[white, very thick] (0,0.6) rectangle (0,0.7);
        \draw[black, very thick, fill=lightgray] (0.2,0) rectangle (0.4,0.2);
        \draw[black, very thick, fill=lightgray] (0.4,0) rectangle (0.6,0.2);
        \draw[black, very thick, fill=lightgray] (0,0.2) rectangle (0.2,0.4);
        \draw[black, very thick, fill=lightgray] (0,0.4) rectangle (0.2,0.6);
        \draw[white, very thick] (0,-0.1) rectangle (0,0.0);
        \end{tikzpicture} &
        \begin{tikzpicture}[baseline=0.5em]
        \draw[white, very thick] (0,0.6) rectangle (0,0.7);
        \draw[black, very thick, fill=lightgray] (0.4,0) rectangle (0.6,0.2);
        \draw[black, very thick, fill=lightgray] (0.4,0.2) rectangle (0.6,0.4);
        \draw[black, very thick, fill=lightgray] (0,0.4) rectangle (0.2,0.6);
        \draw[black, very thick, fill=lightgray] (0.2,0.4) rectangle (0.4,0.6);
        \draw[white, very thick] (0,-0.1) rectangle (0,0.0);
        \end{tikzpicture}\\ \hline
        
        $[[72,4,8]]$ & 
        $[36, 4, 18]$ & 
        $0.06$ & 
        $1.00$ & 
        \begin{tikzpicture}[baseline=0.5em]
        \draw[white, very thick] (0,0.4) rectangle (0,0.5);
        \draw[black, very thick, fill=lightgray] (0,0) rectangle (0.2,0.2);
        \draw[black, very thick, fill=lightgray] (0.2,0) rectangle (0.4,0.2);
        \draw[black, very thick, fill=lightgray] (0,0.2) rectangle (0.2,0.4);
        \draw[white, very thick] (0,-0.1) rectangle (0,0.0);
        \end{tikzpicture} &
        \begin{tikzpicture}[baseline=0.5em]
        \draw[white, very thick] (0,0.4) rectangle (0,0.5);
        \draw[black, very thick, fill=lightgray] (0.2,0) rectangle (0.4,0.2);
        \draw[black, very thick, fill=lightgray] (0,0.2) rectangle (0.2,0.4);
        \draw[black, very thick, fill=lightgray] (0.2,0.2) rectangle (0.4,0.4);
        \draw[white, very thick] (0,-0.1) rectangle (0,0.0);
        \end{tikzpicture}\\ \hline
        
        \cellcolor{orange!10}{$[[162,10,6]]$} & 
        \cellcolor{orange!10}{$[81, 5, 27]$} & 
        \cellcolor{orange!10}{$0.06$} & 
        \cellcolor{orange!10}{$\bm{1.67}$} & 
        \cellcolor{orange!10}{\begin{tikzpicture}[baseline=0.5em]
        \draw[orange!10, very thick] (0,0.6) rectangle (0,0.7);
        \draw[black, very thick, fill=lightgray] (0,0) rectangle (0.2,0.2);
        \draw[black, very thick, fill=lightgray] (0.2,0.2) rectangle (0.4,0.4);
        \draw[black, very thick, fill=lightgray] (0.4,0.2) rectangle (0.6,0.4);
        \draw[black, very thick, fill=lightgray] (0,0.4) rectangle (0.2,0.6);
        \draw[orange!10, very thick] (0,-0.1) rectangle (0,0.0);
        \end{tikzpicture}} &
        \cellcolor{orange!10}{\begin{tikzpicture}[baseline=0.5em]
        \draw[orange!10, very thick] (0,0.6) rectangle (0,0.7);
        \draw[black, very thick, fill=lightgray] (0.4,0) rectangle (0.6,0.2);
        \draw[black, very thick, fill=lightgray] (0,0.2) rectangle (0.2,0.4);
        \draw[black, very thick, fill=lightgray] (0.2,0.2) rectangle (0.4,0.4);
        \draw[black, very thick, fill=lightgray] (0.4,0.4) rectangle (0.6,0.6);
        \draw[orange!10, very thick] (0,-0.1) rectangle (0,0.0);
        \end{tikzpicture}}\\ \hline
        
        $[[162,22,6]]$ & 
        $[81, 11, 9]$ & 
        $0.14$ & 
        $1.22$ & 
        \begin{tikzpicture}[baseline=0.5em]
        \draw[white, very thick] (0,0.6) rectangle (0,0.7);
        \draw[black, very thick, fill=lightgray] (0.2,0) rectangle (0.4,0.2);
        \draw[black, very thick, fill=lightgray] (0.4,0) rectangle (0.6,0.2);
        \draw[black, very thick, fill=lightgray] (0,0.2) rectangle (0.2,0.4);
        \draw[black, very thick, fill=lightgray] (0,0.4) rectangle (0.2,0.6);
        \draw[white, very thick] (0,-0.1) rectangle (0,0.0);
        \end{tikzpicture} &
        \begin{tikzpicture}[baseline=0.5em]
        \draw[white, very thick] (0,0.6) rectangle (0,0.7);
        \draw[black, very thick, fill=lightgray] (0.4,0) rectangle (0.6,0.2);
        \draw[black, very thick, fill=lightgray] (0.4,0.2) rectangle (0.6,0.4);
        \draw[black, very thick, fill=lightgray] (0,0.4) rectangle (0.2,0.6);
        \draw[black, very thick, fill=lightgray] (0.2,0.4) rectangle (0.4,0.6);
        \draw[white, very thick] (0,-0.1) rectangle (0,0.0);
        \end{tikzpicture}\\ \hline
        
        \cellcolor{orange!10}{$[[288,16,8]]$} & 
        \cellcolor{orange!10}{$[144, 20, 42]$} & 
        \cellcolor{orange!10}{$0.06$} & 
        \cellcolor{orange!10}{$\bm{2.33}$}  & 
        \cellcolor{orange!10}{\begin{tikzpicture}[baseline=0.5em]
        \draw[orange!10, very thick] (0,0.6) rectangle (0,0.7);
        \draw[black, very thick, fill=lightgray] (0,0) rectangle (0.2,0.2);
        \draw[black, very thick, fill=lightgray] (0.2,0.2) rectangle (0.4,0.4);
        \draw[black, very thick, fill=lightgray] (0.4,0.2) rectangle (0.6,0.4);
        \draw[black, very thick, fill=lightgray] (0,0.4) rectangle (0.2,0.6);
        \draw[orange!10, very thick] (0,-0.1) rectangle (0,0.0);
        \end{tikzpicture}} &
        \cellcolor{orange!10}{\begin{tikzpicture}[baseline=0.5em]
        \draw[orange!10, very thick] (0,0.6) rectangle (0,0.7);
        \draw[black, very thick, fill=lightgray] (0.4,0) rectangle (0.6,0.2);
        \draw[black, very thick, fill=lightgray] (0,0.2) rectangle (0.2,0.4);
        \draw[black, very thick, fill=lightgray] (0.2,0.2) rectangle (0.4,0.4);
        \draw[black, very thick, fill=lightgray] (0.4,0.4) rectangle (0.6,0.6);
        \draw[orange!10, very thick] (0,-0.1) rectangle (0,0.0);
        \end{tikzpicture}}\\ \hline
        
        \cellcolor{orange!10}{$[[288,12,12]]$} & 
        \cellcolor{orange!10}{$[144, 7, 54]$} & 
        \cellcolor{orange!10}{$0.04$} & 
        \cellcolor{orange!10}{$\bm{2.25}$} & 
        \cellcolor{orange!10}{\begin{tikzpicture}[baseline=0.5em]
        \draw[orange!10, very thick] (0,0.6) rectangle (0,0.7);
        \draw[black, very thick, fill=lightgray] (0.2,0) rectangle (0.4,0.2);
        \draw[black, very thick, fill=lightgray] (0.2,0.2) rectangle (0.4,0.4);
        \draw[black, very thick, fill=lightgray] (0.4,0.2) rectangle (0.6,0.4);
        \draw[black, very thick, fill=lightgray] (0,0.4) rectangle (0.2,0.6);
        \draw[orange!10, very thick] (0,-0.1) rectangle (0,0.0);
        \end{tikzpicture}} &
        \cellcolor{orange!10}{
        \begin{tikzpicture}[baseline=0.5em]
        \draw[orange!10, very thick] (0,0.6) rectangle (0,0.7);
        \draw[black, very thick, fill=lightgray] (0.4,0) rectangle (0.6,0.2);
        \draw[black, very thick, fill=lightgray] (0,0.2) rectangle (0.2,0.4);
        \draw[black, very thick, fill=lightgray] (0.2,0.2) rectangle (0.4,0.4);
        \draw[black, very thick, fill=lightgray] (0.2,0.4) rectangle (0.4,0.6);
        \draw[orange!10, very thick] (0,-0.1) rectangle (0,0.0);
        \end{tikzpicture}}\\ \hline
        
        \cellcolor{orange!10}{$[[288,22,6]]$} & 
        \cellcolor{orange!10}{$[144, 15, 24]$} & 
        \cellcolor{orange!10}{$0.08$} & 
        \cellcolor{orange!10}{$\bm{1.83}$} &  
        \cellcolor{orange!10}{\begin{tikzpicture}[baseline=0.5em]
        \draw[orange!10, very thick] (0,0.8) rectangle (0,0.9);
        \draw[black, very thick, fill=lightgray] (0,0) rectangle (0.2,0.2);
        \draw[black, very thick, fill=lightgray] (0.6,0.2) rectangle (0.8,0.4);
        \draw[black, very thick, fill=lightgray] (0.2,0.4) rectangle (0.4,0.6);
         \draw[black, very thick, fill=lightgray] (0,0.6) rectangle (0.2,0.8);
        \draw[orange!10, very thick] (0,-0.1) rectangle (0,0.0);
        \end{tikzpicture}} &
        \cellcolor{orange!10}{\begin{tikzpicture}[baseline=0.5em]
        \draw[orange!10, very thick] (0,0.8) rectangle (0,0.9);
        \draw[black, very thick, fill=lightgray] (0.6,0) rectangle (0.8,0.2);
        \draw[black, very thick, fill=lightgray] (0.4,0.2) rectangle (0.6,0.4);
        \draw[black, very thick, fill=lightgray] (0,0.4) rectangle (0.2,0.6);
        \draw[black, very thick, fill=lightgray] (0.6,0.6) rectangle (0.8,0.8);
        \draw[orange!10, very thick] (0,-0.1) rectangle (0,0.0);
        \end{tikzpicture}}\\ \hline
        
        $[[288,14,6]]$ & 
        $[144, 7, 36]$ & 
        $0.05$ & 
        $1.75$ & 
        \begin{tikzpicture}[baseline=0.5em]
        \draw[white, very thick] (0,0.8) rectangle (0,0.9);
        \draw[black, very thick, fill=lightgray] (0.2,0) rectangle (0.4,0.2);
        \draw[black, very thick, fill=lightgray] (0.6,0.2) rectangle (0.8,0.4);
        \draw[black, very thick, fill=lightgray] (0,0.4) rectangle (0.2,0.6);
         \draw[black, very thick, fill=lightgray] (0,0.6) rectangle (0.2,0.8);
        \draw[white, very thick] (0,-0.1) rectangle (0,0.0);
        \end{tikzpicture} &
        \begin{tikzpicture}[baseline=0.5em]
        \draw[white, very thick] (0,0.8) rectangle (0,0.9);
        \draw[black, very thick, fill=lightgray] (0.6,0) rectangle (0.8,0.2);
        \draw[black, very thick, fill=lightgray] (0.6,0.2) rectangle (0.8,0.4);
        \draw[black, very thick, fill=lightgray] (0,0.4) rectangle (0.2,0.6);
        \draw[black, very thick, fill=lightgray] (0.4,0.6) rectangle (0.6,0.8);
        \draw[white, very thick] (0,-0.1) rectangle (0,0.0);
        \end{tikzpicture}\\ \hline
        
        $[[288,30,6]]$ & 
        $[144, 15, 12]$ & 
        $0.10$ & 
        $1.25$ & 
        \begin{tikzpicture}[baseline=0.5em]
        \draw[white, very thick] (0,0.8) rectangle (0,0.9);
        \draw[black, very thick, fill=lightgray] (0.6,0) rectangle (0.8,0.2);
        \draw[black, very thick, fill=lightgray] (0.2,0.2) rectangle (0.4,0.4);
        \draw[black, very thick, fill=lightgray] (0,0.4) rectangle (0.2,0.6);
         \draw[black, very thick, fill=lightgray] (0,0.6) rectangle (0.2,0.8);
        \draw[white, very thick] (0,-0.1) rectangle (0,0.0);
        \end{tikzpicture} &
        \begin{tikzpicture}[baseline=0.5em]
        \draw[white, very thick] (0,0.8) rectangle (0,0.9);
        \draw[black, very thick, fill=lightgray] (0.6,0) rectangle (0.8,0.2);
        \draw[black, very thick, fill=lightgray] (0.6,0.2) rectangle (0.8,0.4);
        \draw[black, very thick, fill=lightgray] (0.4,0.4) rectangle (0.6,0.6);
         \draw[black, very thick, fill=lightgray] (0,0.6) rectangle (0.2,0.8);
        \draw[white, very thick] (0,-0.1) rectangle (0,0.0);
        \end{tikzpicture} \\ \hline
        
        \cellcolor{orange!10}{$[[450,22,6]]$} & 
        \cellcolor{orange!10}{$[225, 27, 45 %51
        ]$} & 
        \cellcolor{orange!10}{$0.05$} & 
        \cellcolor{orange!10}{$\bm{2.20}$} & 
        \cellcolor{orange!10}{\begin{tikzpicture}[baseline=0.5em]
        \draw[orange!10, very thick] (0,0.8) rectangle (0,0.9);
        \draw[black, very thick, fill=lightgray] (0,0) rectangle (0.2,0.2);
        \draw[black, very thick, fill=lightgray] (0.6,0.2) rectangle (0.8,0.4);
        \draw[black, very thick, fill=lightgray] (0.4,0.4) rectangle (0.6,0.6);
         \draw[black, very thick, fill=lightgray] (0,0.6) rectangle (0.2,0.8);
        \draw[orange!10, very thick] (0,-0.1) rectangle (0,0.0);
        \end{tikzpicture}} &
        \cellcolor{orange!10}{\begin{tikzpicture}[baseline=0.5em]
        \draw[orange!10, very thick] (0,0.8) rectangle (0,0.9);
        \draw[black, very thick, fill=lightgray] (0.6,0) rectangle (0.8,0.2);
        \draw[black, very thick, fill=lightgray] (0.2,0.2) rectangle (0.4,0.4);
        \draw[black, very thick, fill=lightgray] (0,0.4) rectangle (0.2,0.6);
        \draw[black, very thick, fill=lightgray] (0.6,0.6) rectangle (0.8,0.8);
        \draw[orange!10, very thick] (0,-0.1) rectangle (0,0.0);
        \end{tikzpicture}}\\ \hline
        
        $[[450,10,10]]$ & 
        $[225, 13, 75 %85
        ]$ & 
        $0.02$ & 
        $1.67$ & 
        \begin{tikzpicture}[baseline=0.5em]
        \draw[white, very thick] (0,0.8) rectangle (0,0.9);
        \draw[black, very thick, fill=lightgray] (0.2,0) rectangle (0.4,0.2);
        \draw[black, very thick, fill=lightgray] (0.2,0.4) rectangle (0.4,0.6);
        \draw[black, very thick, fill=lightgray] (0.6,0.4) rectangle (0.8,0.6);
        \draw[black, very thick, fill=lightgray] (0,0.6) rectangle (0.2,0.8);
        \draw[white, very thick] (0,-0.1) rectangle (0,0.0);
        \end{tikzpicture} &
        \begin{tikzpicture}[baseline=0.5em]
        \draw[white, very thick] (0,0.8) rectangle (0,0.9);
        \draw[black, very thick, fill=lightgray] (0.6,0) rectangle (0.8,0.2);
        \draw[black, very thick, fill=lightgray] (0,0.2) rectangle (0.2,0.4);
        \draw[black, very thick, fill=lightgray] (0.4,0.2) rectangle (0.6,0.4);
        \draw[black, very thick, fill=lightgray] (0.4,0.6) rectangle (0.6,0.8);
        \draw[white, very thick] (0,-0.1) rectangle (0,0.0);
        \end{tikzpicture} \\ \hline
        
        $[[450,18,10]]$ & 
        $[225, 25, 45]$ & 
        $0.04$ & 
        $1.80$ & 
        \begin{tikzpicture}[baseline=0.5em]
        \draw[white, very thick] (0,0.8) rectangle (0,0.9);
        \draw[black, very thick, fill=lightgray] (0.4,0) rectangle (0.6,0.2);
        \draw[black, very thick, fill=lightgray] (0.2,0.4) rectangle (0.4,0.6);
        \draw[black, very thick, fill=lightgray] (0.6,0.4) rectangle (0.8,0.6);
        \draw[black, very thick, fill=lightgray] (0,0.6) rectangle (0.2,0.8);
        \draw[white, very thick] (0,-0.1) rectangle (0,0.0);
        \end{tikzpicture} &
        \begin{tikzpicture}[baseline=0.5em]
        \draw[white, very thick] (0,0.8) rectangle (0,0.9);
        \draw[black, very thick, fill=lightgray] (0.6,0) rectangle (0.8,0.2);
        \draw[black, very thick, fill=lightgray] (0,0.2) rectangle (0.2,0.4);
        \draw[black, very thick, fill=lightgray] (0.4,0.2) rectangle (0.6,0.4);
        \draw[black, very thick, fill=lightgray] (0.2,0.6) rectangle (0.4,0.8);
        \draw[white, very thick] (0,-0.1) rectangle (0,0.0);
        \end{tikzpicture} \\ \hline
        
        $[[450,30,8]]$ & 
        $[225, 27, 15]$ & 
        $0.07$ & 
        $1.00$ & 
        \begin{tikzpicture}[baseline=0.5em]
        \draw[white, very thick] (0,0.8) rectangle (0,0.9);
        \draw[black, very thick, fill=lightgray] (0.6,0) rectangle (0.8,0.2);
        \draw[black, very thick, fill=lightgray] (0.2,0.2) rectangle (0.4,0.4);
        \draw[black, very thick, fill=lightgray] (0,0.4) rectangle (0.2,0.6);
        \draw[black, very thick, fill=lightgray] (0,0.6) rectangle (0.2,0.8);
        \draw[white, very thick] (0,-0.1) rectangle (0,0.0);
        \end{tikzpicture} &
        \begin{tikzpicture}[baseline=0.5em]
        \draw[white, very thick] (0,0.8) rectangle (0,0.9);
        \draw[black, very thick, fill=lightgray] (0.6,0) rectangle (0.8,0.2);
        \draw[black, very thick, fill=lightgray] (0.6,0.2) rectangle (0.8,0.4);
        \draw[black, very thick, fill=lightgray] (0.4,0.4) rectangle (0.6,0.6);
        \draw[black, very thick, fill=lightgray] (0,0.6) rectangle (0.2,0.8);
        \draw[white, very thick] (0,-0.1) rectangle (0,0.0);
        \end{tikzpicture} \\ \hline
        
    \end{tabular}
\end{table}

\subsection{Code families} 

In order to find the most promising stabilizer shapes we constrained the honeycomb lattice to have the same height and width. We found codes with interesting parameters in \cref{tab:Romanesco codes}. Here we vary the height and width for each stabilizer shape to find the code families with the highest $v_\infty=kd_c/n$ ratio. We define a family by fixing the number of encoded qubits $k$ and by making sure that all instances in the family have the same functional for the classical distance, $d_c = f(n)$, where $f(n)$ is a function of the number of data qubits $n$. As a result, all members of the family have the same ratio $v_\infty$. Each family is defined by a set of geometric constraints on the lattice. We consider lattices with less than 1000 qubits for the search but remark that there might be interesting families for larger lattice code sizes, particularly for $m=4$, that could result in higher encoding rates. The codes in \cref{tab:Romanesco codes} are not necessarily part of these families. The code families are reported in \cref{tab:Romanesco codes rectangular}.

\section{Sampling logical representatives}\label{app:logical representatives}

In this appendix, we present the approach we used to sample logical representatives of the Romanesco codes with different ratios of X and Z Pauli operators, in more detail. We use \cref{algo:representatives} to sample the logical representatives, with different values of depolarizing rates and noise biases. We used all combinations of depolarizing rates $p \in \{10^{-4}, 10^{-3}, 10^{-2}, 10^{-1}, 0.49\}$ and noise biases $\eta \in \{1, 10^1, 10^3, 10^5, 10^7, 10^9, 10^{11}, 10^{13}\}$. Importantly, we also apply a Hadamard rotation on all gray qubits in the noise channels so that we get the minimum-weight representatives of the non-CSS code while using the parity check matrices of the CSS code. This simplifies our task because with a CSS code we can use the X parity check matrix to get random X logical representatives that can then be used to find minimum-weight Z logical representatives s by solving the decoding problem with the Z parity check matrix. See \cref{app:rotating decoding problem} for an explanation of why it is equivalent. In the infinite bias limit, the Z noise channel then has depolarizing noise only on black qubits whereas the X noise channel has depolarizing noise only on the gray qubits. Since the X and Z parity check matrices are equivalent, we focus here only on generating minimum-weight Z logical representatives.

\begin{algorithm}[h!]
    \caption{
    We solve a modified decoding problem to obtain a list of minimum-weight logical representatives for any CSS code with parity check matrices $H_X$ and $H_Z$ with different numbers of X Pauli operators. The roles of $H_X$ and $H_Z$ can be swapped. Here we use the BP+OSD-CS decoder with the BP and OSD-CS parameters as in \cref{algo:code distance}.}\label{algo:representatives}
    \begin{algorithmic}
        \Procedure{Z Logicals}{$H_X$, $H_Z$, $[p_{Xi}]$, $[p_{Zi}]$}
        \State \textbf{input: } X parity check matrix $H_X$, Z parity check matrix $H_Z$, X depolarizing rate per qubit $[p_{Xi}]$, Z depolarizing rate per qubit $[p_{Zi}]$
        \State \textbf{output: } List of Z logicals
        \State $[L_Z] \gets$ list of Z logicals
        \State // \textit{Get test X logicals}
        \State $[L_X] \gets$ list of X test logicals
        \For{$L_X \in \mathrm{ker}_{\mathrm{mod}2}(H_Z)$}
        \State $\tilde{H}_X = \mathrm{vstack}(H_X, L_X)$
        \State // \textit{check it is not in the rowspace of $H_X$}
        \If{$\mathrm{rk}_{\mathrm{mod}2}(\tilde{H}_X)\neq \mathrm{rk}_{\mathrm{mod}2}(H_X)$}
        \State add $L_X$ to $[L_X]$
        \EndIf
        \EndFor
        \State // \textit{Find low-weight Z logicals}
        \State // $[\tilde{L}_X] \gets$ \textit{random combinations of $[L_X]$}
        \For{$\tilde{L}_X \in [\tilde{L}_X]$}
        \State // \textit{set up the decoder}
        \State $D_X = \mathrm{BPOSD}(\mathrm{vstack}(H_X,\tilde{L}_X), [p_{Xi}])$
        \State $s_X = \mathrm{vstack}(\vec{0}_{\mathrm{nrow}(H_X)}, 1)$ // syndrome
        \State $L_Z = D_X(s_X)$ // \textit{decode to get Z logical}
        \State add $L_Z$ to $[L_Z]$
        \EndFor
        \Return $[L_Z]$
        \EndProcedure
    \end{algorithmic}
\end{algorithm}

In \cref{tab:mixed logicals} we show the minimum weight of a logical representative with $s$ Pauli X operators for each code in \cref{tab:Romanesco codes} using \cref{algo:representatives}. As explained in the main text, if the quantum code has distance $d$ then we only need to consider $0\leq s\leq \lceil d/2\rceil$. Indeed, this is because the minimum-weight representative in the quantum code has weight $d$ and is made of half $X$ and half $Z$ Pauli operators.

\begin{table}[h!]
    \centering
    \caption{Distance $d_s$ of the lowest-weight logical representative with $s$ non-Z Pauli operators for the periodic codes in \cref{tab:Romanesco codes}. The saturation is proportional to the drop in distance relative to $s=0$.}
    \label{tab:mixed logicals}
    \begin{tabular}{c c c c c c c c}
        %\hline
        & \multicolumn{7}{c}{\cellcolor{black!10}{$\bm{s}$}} \\
        %\hline
        \cellcolor{black!10}{$\bm{[[n,k,d]]}$} & \cellcolor{black!10}{$\bm{0}$} & \cellcolor{black!10}{$\bm{1}$} & \cellcolor{black!10}{$\bm{2}$} & \cellcolor{black!10}{$\bm{3}$} & \cellcolor{black!10}{$\bm{4}$} & \cellcolor{black!10}{$\bm{5}$} & \cellcolor{black!10}{$\bm{6}$} \\ %\hline
        
        \cellcolor{black!10}{$[[72,12,4]]$} & \cellcolor{orange!10}{12} & \cellcolor{orange!10}{$\varnothing$} & \cellcolor{orange!90}{4} & \cellcolor{black!75} & \cellcolor{black!75} & \cellcolor{black!75} & \cellcolor{black!75} \\ %\hline
        
        \cellcolor{black!10}{$[[72,16,4]]$} & \cellcolor{orange!10}{6} & \cellcolor{orange!10}{$\varnothing$} & \cellcolor{orange!90}{4} & \cellcolor{black!75} & \cellcolor{black!75} & \cellcolor{black!75} & \cellcolor{black!75} \\ %\hline
        
        \cellcolor{black!10}{$[[72,4,8]]$} & \cellcolor{orange!10}{18} & \cellcolor{orange!10}{$\varnothing$} & \cellcolor{orange!60}{12} & \cellcolor{orange!75}{10} & \cellcolor{orange!90}{8} & \cellcolor{black!75} & \cellcolor{black!75} \\ %\hline
        
        \cellcolor{black!10}{$[[162,10,6]]$} & \cellcolor{orange!10}{27} & \cellcolor{orange!10}{30} & \cellcolor{orange!30} {25} & \cellcolor{orange!90}{6} & \cellcolor{black!75} & \cellcolor{black!75} & \cellcolor{black!75} \\ %\hline
        
        \cellcolor{black!10}{$[[162,22,6]]$} & \cellcolor{orange!10}{9} & \cellcolor{orange!10} 21 & \cellcolor{orange!10} 23 & \cellcolor{orange!90}{6} & \cellcolor{black!75} & \cellcolor{black!75} & \cellcolor{black!75} \\ %\hline
        
        \cellcolor{black!10}{$[[288,16,8]]$} & \cellcolor{orange!10}{42} & \cellcolor{orange!10}{$\varnothing$} & \cellcolor{orange!10}{$\varnothing$} & \cellcolor{orange!25}{40} & \cellcolor{orange!90}{8} & \cellcolor{black!75} & \cellcolor{black!75} \\ %\hline
        
        \cellcolor{black!10}{$[[288,12,12]]$} & \cellcolor{orange!10}{54} & \cellcolor{orange!10}{$\varnothing$} & \cellcolor{orange!30}{36} & \cellcolor{orange!30} {$\varnothing$} & \cellcolor{orange!60}{24} & \cellcolor{orange!60}{$\varnothing$} & \cellcolor{orange!90}{12} \\ %\hline
        
        \cellcolor{black!10}{$[[288,22,6]]$} & \cellcolor{orange!10}{24} & \cellcolor{orange!10}{$\varnothing$} & \cellcolor{orange!10}{26} & \cellcolor{orange!90}{6} & \cellcolor{black!75} & \cellcolor{black!75} & \cellcolor{black!75} \\ %\hline
        
        \cellcolor{black!10}{$[[288,14,6]]$} & \cellcolor{orange!10}{36} & \cellcolor{orange!10}{44} & \cellcolor{orange!10}{ 40} & \cellcolor{orange!90}{6} & \cellcolor{black!75} & \cellcolor{black!75} & \cellcolor{black!75} \\ %\hline
        
        \cellcolor{black!10}{$[[288,30,6]]$} & \cellcolor{orange!10}{12} & \cellcolor{orange!10}{20} & \cellcolor{orange!10}{16} & \cellcolor{orange!90}{6} & \cellcolor{black!75} & \cellcolor{black!75} & \cellcolor{black!75} \\ %\hline
        
        \cellcolor{black!10}{$[[450,22,6]]$} & \cellcolor{orange!10}{45} & \cellcolor{orange!10}{$\varnothing$} &\cellcolor{orange!10}{$\varnothing$} & \cellcolor{orange!90}{6} & \cellcolor{black!75} & \cellcolor{black!75} & \cellcolor{black!75} \\ %\hline
        
        \cellcolor{black!10}{$[[450,10,10]]$} & \cellcolor{orange!10}{75} & \cellcolor{orange!10}{$\varnothing$} & \cellcolor{orange!10}{86} & \cellcolor{orange!10}{74} & \cellcolor{orange!30}{64} & \cellcolor{orange!90}{10} & \cellcolor{black!75} \\ %\hline
        
        \cellcolor{black!10}{$[[450,18,10]]$} & \cellcolor{orange!10}{45} & \cellcolor{orange!10}{$\varnothing$} & \cellcolor{orange!10}{$\varnothing$} & \cellcolor{orange!10}{57} & \cellcolor{orange!10}{58} & \cellcolor{orange!90}{10} & \cellcolor{black!75} \\ %\hline
        
        \cellcolor{black!10}{$[[450,30,8]]$} & \cellcolor{orange!10}{15} & \cellcolor{orange!10}{$\varnothing$} &  \cellcolor{orange!10}{$\varnothing$} & \cellcolor{orange!10}{32} & \cellcolor{orange!90}{8} & \cellcolor{black!75} & \cellcolor{black!75} \\ %\hline
        
    \end{tabular}
\end{table}

\begin{figure}[h!]
    \centering
    \includegraphics[width=\linewidth]{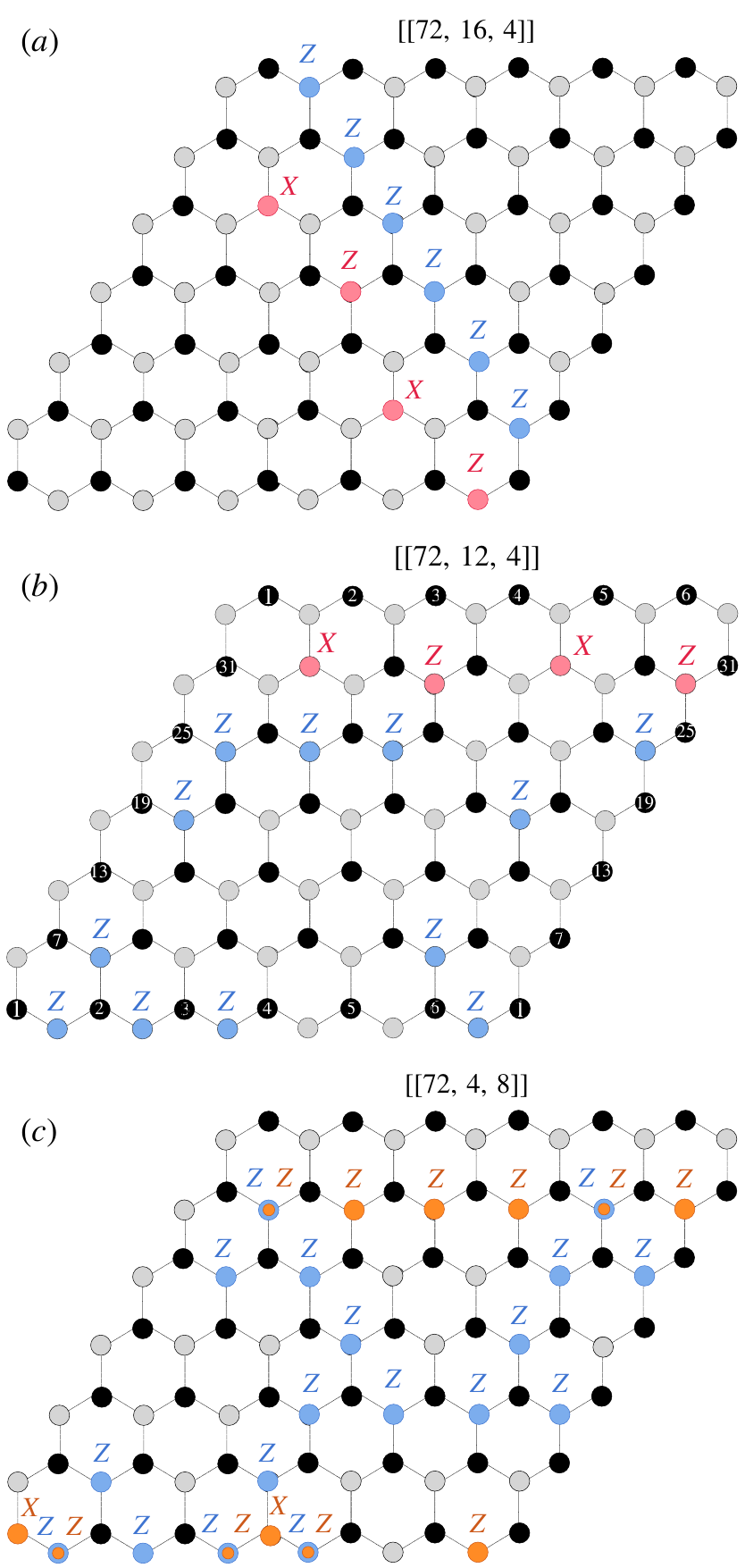}
    \caption{The first two lowest-weight mixed logical operators for the 72 qubits codes. The Z Pauli operators are all on the gray qubits here but could also instead be on black qubits by symmetry whereas the X Pauli operators are on the black qubits. The Romanesco codes are advantageous for noise bias as long as the string-like operators of distance $d$ (with $d/2$ Z and $d/2$ non-Z Pauli operators) have a low contribution to the logical error rate.}
    \label{fig:72 mixed logicals app}
\end{figure}

In \cref{fig:72 mixed logicals app} we showed two logical representatives of the $[[72,12,4]]$, $[[72,16,4]]$ and $[[72,4,8]]$ codes. In all three codes, we observe string-like logical operators that are mixed (half X and half Z) with weight $d$ (not shown for $[[72,4,8]]$ for conciseness). We observe fractal-like all-Z representatives for $[[72,12,4]]$ and $[[72,4,8]]$. The all-Z representatives in $[[72,16,4]$ are however string-like. The codes reported in \cref{tab:Romanesco codes} do not necessarily share similar properties. This can be appreciated from their different profiles in \cref{tab:mixed logicals}. However, all the members of a code family in \cref{tab:Romanesco codes rectangular} will be similar as observed in \cref{tab:mixed logicals families}.

\section{Decoding the non-CSS code}\label{app:rotating decoding problem}

There are two ways to simulate the non-CSS code. A first approach would be to directly rotate the parity check matrix with the Hadamard rotation $U_q$. Alternatively, we can rotate the noise channels, which is what we did here. It is possible to solve the decoding problem for the non-CSS code using the CSS version of the code (without the unitary rotation on the gray qubits) and instead rotating the noise channels. Let us define the error as 
\begin{equation}
    e = ( (e_{Z1}, e_{Z2}) | (e_{X1}, e_{X2}))^T,
\end{equation}
where $e_{Pi}$ is a $P=X,Z$-type error on sector $i$ of the data qubits with $i=1$ for black qubits and $i=2$ for gray qubits. The syndrome in the non-CSS code $C_\clubsuit$ with parity check matrix $H^\star$ is then 
\begin{equation}
    s^\star = H^\star  \cdot e = (U_q^\dagger \cdot H  \cdot U_q) \cdot e = U_q^\dagger \cdot H\cdot (U_q \cdot e),
\end{equation}
where $H$ is the parity check matrix of the CSS code $C$,
\begin{equation}
    H = \begin{pmatrix}H_X & 0 \\ 0 & H_Z\end{pmatrix}.
\end{equation}
We can therefore solve the decoding problem with the rotated error $U_q \cdot e$ using the CSS code parity check matrix $H$. Here $U_q \cdot e$ can be directly generated by rotating the noise channels, i.e. $p_X \to p_Z$ and $p_Z\to p_X$ on each gray qubit.

\section{Boundary conditions}\label{app: boundary conditions}

In this appendix, we look at Romanesco codes defined on the cylinder manifold (with quasi-periodic boundary conditions) and the two-dimensional plane (with fully open boundary conditions). We show the stabilizers of the codes on the different manifolds, define code families and look at the weight of the mixed logical representatives with different ratios of X and Z Pauli operators.

\subsection{Stabilizer generators}

\begin{figure}[h!]
    \centering
    \includegraphics[width=\linewidth]{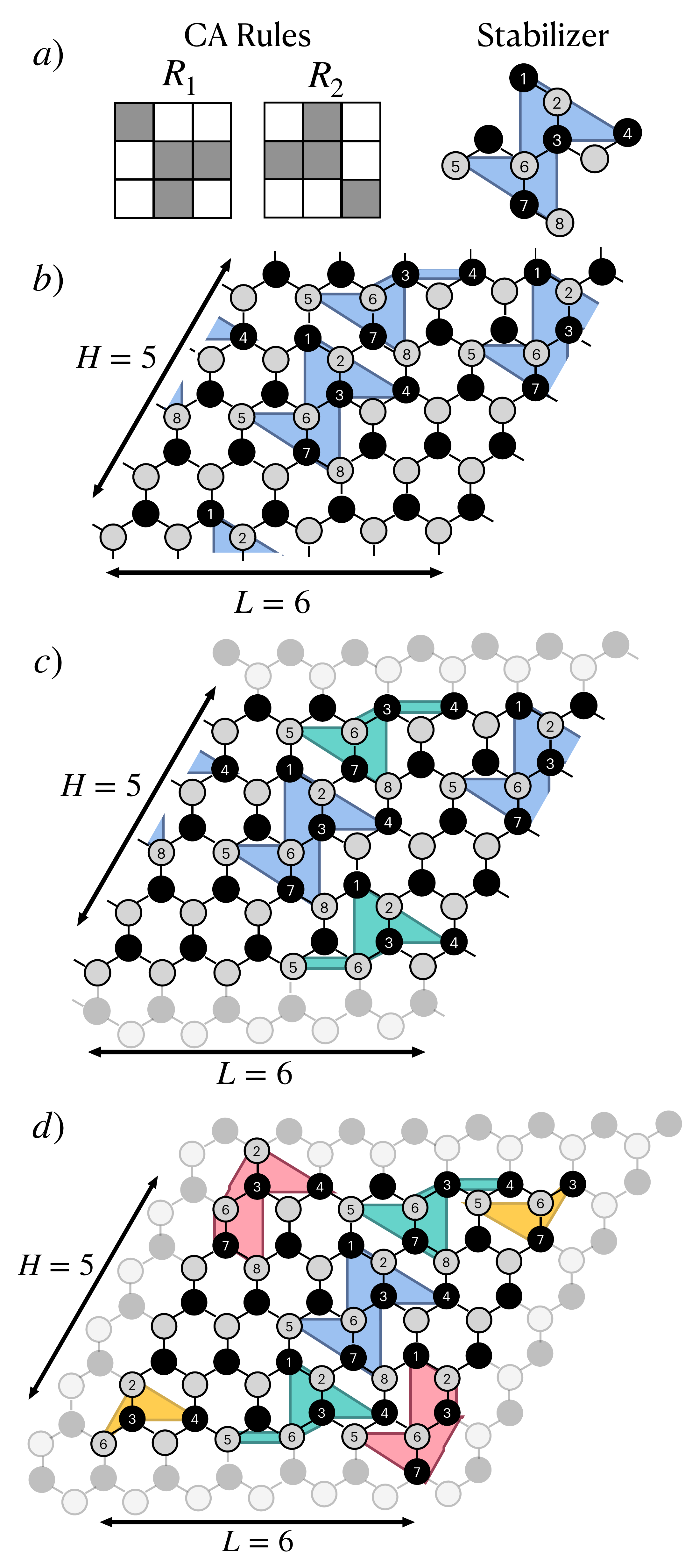}
    \caption{a) Cellular automaton rules $R_1$ and $R_2$ and the resulting stabilizer for the quantum code on the torus. b) Stabilizers of the code family $[N,12,D_{1,1}]]$ in \cref{tab:Romanesco codes rectangular} on the torus. c) Stabilizers of the code family $[N^\prime,6,D_{1,1}]]$ in \cref{tab:Romanesco codes cylinder} on a cylinder. d) Stabilizers of the codes in \cref{tab:Romanesco codes 2d plane} on a two-dimensional plane with open boundary conditions.}
    \label{fig:cylinder family 1}
\end{figure}

\begin{figure}[h!]
    \centering
    \includegraphics[width=\linewidth]{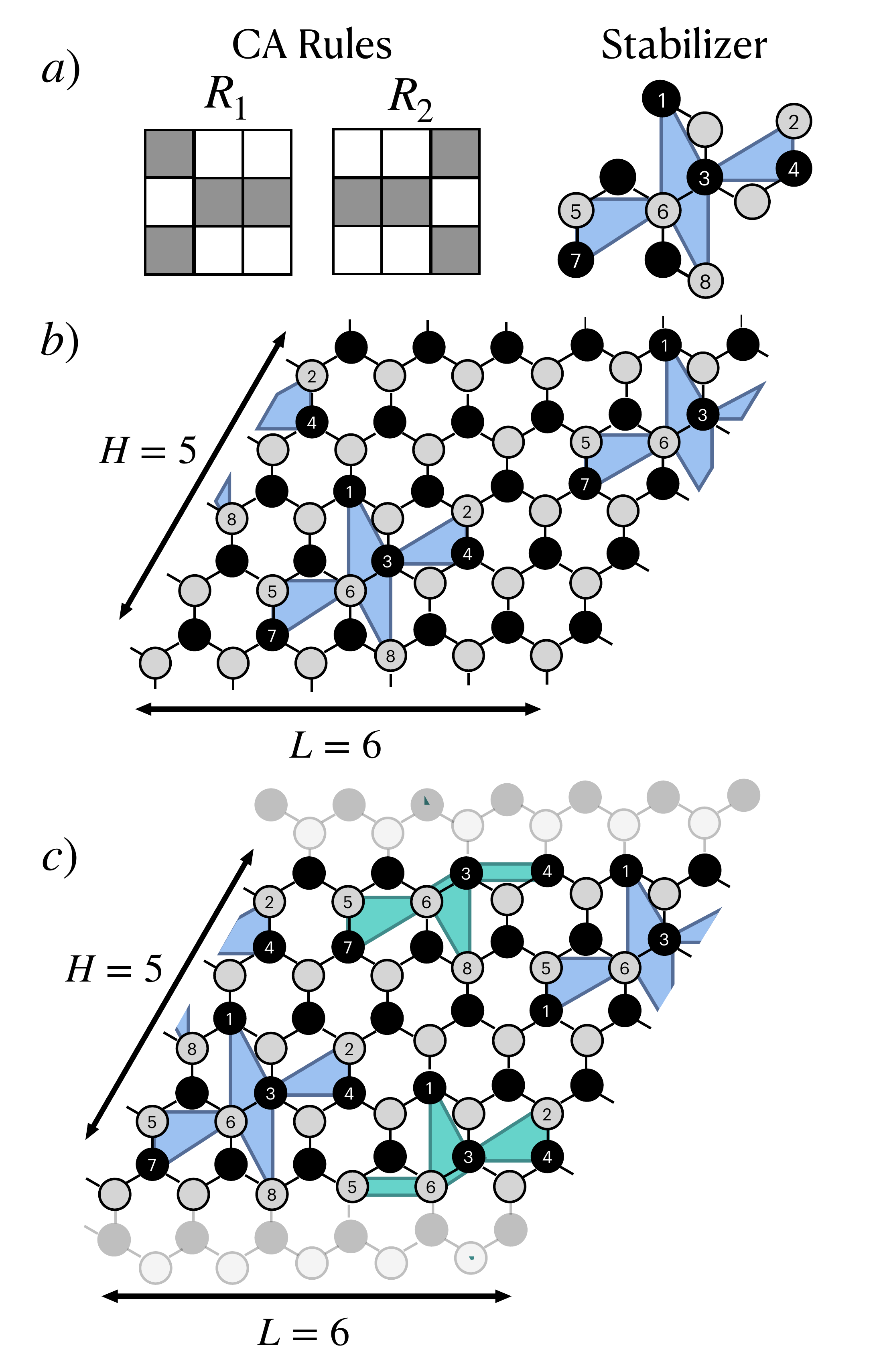}
    \caption{a) Cellular automaton rules $R_1$ and $R_2$ and the resulting stabilizer for the quantum code on the torus. b) Stabilizers of the code family $[N,16,D_{1,2/3}]]$ in \cref{tab:Romanesco codes rectangular} on the torus. c) Stabilizers of the code family $[N^\prime,8,D_{1,2/3}]]$ in \cref{tab:Romanesco codes cylinder} on a cylinder.}
    \label{fig:cylinder family 2}
\end{figure}

In \cref{fig:cylinder family 1,fig:cylinder family 2} we show the stabilizers for the code families in \cref{tab:Romanesco codes rectangular} on the torus, the cylinder and two-dimensional plane. We will refer to the codes with the bulk stabilizer in \cref{fig:cylinder family 1} a) as ``bowtie'' codes and those with the bulk stabilizer in \cref{fig:cylinder family 2} a) as ``butterfly'' codes. We were only able to find bowtie codes on the two-dimensional plane with fully open boundaries. It is possible that the use of different truncated stabilizers at the lateral boundaries could allow us to find butterfly codes in the two-dimensional plane. The approach that we used to find the stabilizers at the boundary is detailed in the main text. We simply truncate the stabilizers at the boundary and ensure that the truncated stabilizers have even weight so that the commutation relations are satisfied.

\subsection{Code families}

The code families on the torus are presented in \cref{tab:Romanesco codes rectangular}. The code families on the cylinder are also shown in \cref{tab:Romanesco codes cylinder}. As highlighted in the main text, the code families on the cylinder have parameters that are similar to those on the torus, with the exception that the number of encoded qubits $k$ is reduced by half. This is similar to what happens when going from the toric code to the surface code as well as for other topological codes. Luckily, we do not see a drop on the code distance $d$ nor in the classical distance $d_c$ despite removing $2L$ data qubits.

Next we move to the two-dimensional plane. We show different examples of bowtie codes on square lattices of size $d-1$ in \cref{tab:Romanesco codes 2d plane extended}. Unfortunately, we see an additional drop in $k$, from 6 on the cylinder to 4 on the two-dimensional plane. It is possible that a different way of opening the lateral boundaries could have preserved the number of encoded qubits.

\begin{table}[t!]
    \centering
    \caption{Examples of distance-$d$ Romanesco codes on a bipartite honeycomb lattice with open boundary conditions, where each sub-lattice lies on a two-dimensional plane with height $H=d-1$ and width $L=d-1$ so that each code has $n = 2HL + 2$ data qubits. We show examples up to 1060 data qubits ($4 \leq d \leq 24$). Here $v_\infty = kd_c/n$. The codes that are highlighted offer the best $v_\infty$ values with $d \geq 8$. However, the logical performance of these codes is better predicted by looking at the minimum weight of mixed logical representatives in \cref{tab:open BC mixed logicals}. All distances here were double-checked with the exact distance calculation function from LDPC~\cite{Roffe_LDPC_Python_tools_2022}.}
    \label{tab:Romanesco codes 2d plane extended}
    \begin{tabular}{c|c}
        %\hline
        % \rowcolor{lightgray}
        \cellcolor{black!10}{$\bm{[[n,k,d]]}_{\square}$} & 
        \cellcolor{black!10}{\textbf{Examples}} 
        % & 
        % \cellcolor{black!10}{$\bm{R_1}$} & 
        % \cellcolor{black!10}{$\bm{R_2}$}
        \\ \hline

        \begin{tabular}{c}
             \hspace{.5em} $[[2(d-1)^2+2, 4, d]]$ \hspace{.5em}\\
             \\
              \begin{tabular}{c|c}
                 \cellcolor{black!10}{\hspace{1em}$\bm{R_1}$\hspace{1em}} & 
        \cellcolor{black!10}{\hspace{1em}$\bm{R_2}$\hspace{1em}} \\
                \hline
                \cellcolor{orange!0}{\begin{tikzpicture}[baseline=0.5em]
                \draw[orange!0, very thick] (0,0.6) rectangle (0,0.7);
                \draw[black, very thick, fill=lightgray] (0.2,0) rectangle (0.4,0.2);
                \draw[black, very thick, fill=lightgray] (0.2,0.2) rectangle (0.4,0.4);
                \draw[black, very thick, fill=lightgray] (0.4,0.2) rectangle (0.6,0.4);
                \draw[black, very thick, fill=lightgray] (0,0.4) rectangle (0.2,0.6);
                \draw[orange!0, very thick] (0,-0.1) rectangle (0,0.0);
                \end{tikzpicture}} &
                \cellcolor{orange!0}{
                \begin{tikzpicture}[baseline=0.5em]
                \draw[orange!0, very thick] (0,0.6) rectangle (0,0.7);
                \draw[black, very thick, fill=lightgray] (0.4,0) rectangle (0.6,0.2);
                \draw[black, very thick, fill=lightgray] (0,0.2) rectangle (0.2,0.4);
                \draw[black, very thick, fill=lightgray] (0.2,0.2) rectangle (0.4,0.4);
                \draw[black, very thick, fill=lightgray] (0.2,0.4) rectangle (0.4,0.6);
                \draw[orange!0, very thick] (0,-0.1) rectangle (0,0.0);
                \end{tikzpicture}}\\
              \end{tabular}
        \end{tabular} & 
        
        \begin{tabular}{c}
        \\[-.7em]
        \begin{tabular}{c | c | c}
         \cellcolor{black!10}{\hspace{.3em} $\bm{d}$ \hspace{.3em}} & \cellcolor{black!10}{\hspace{.3em} $\bm{d_c}$ \hspace{.3em}} & \cellcolor{black!10}{\hspace{.3em} $\bm{v_\infty}$ \hspace{.3em}} \\
         \hline
         4 & 4 & 0.80 \\
         \cellcolor{orange!60}{8} & \cellcolor{orange!60}{16} & \cellcolor{orange!60}{0.64} \\
         \cellcolor{orange!60}{12} & \cellcolor{orange!60}{40} & \cellcolor{orange!60}{0.66} \\
         \cellcolor{orange!15}{16} & \cellcolor{orange!15}{36} & \cellcolor{orange!15}{0.32} \\
         20 & 40 & 0.22 \\
         \cellcolor{orange!60}{24} & \cellcolor{orange!60}{160} & \cellcolor{orange!60}{0.60}\\
        \end{tabular}
        \\
        \\[-.7em]
        \end{tabular}
        % & 
        % \cellcolor{orange!0}{\begin{tikzpicture}[baseline=0.5em]
        % \draw[orange!0, very thick] (0,0.6) rectangle (0,0.7);
        % \draw[black, very thick, fill=lightgray] (0.2,0) rectangle (0.4,0.2);
        % \draw[black, very thick, fill=lightgray] (0.2,0.2) rectangle (0.4,0.4);
        % \draw[black, very thick, fill=lightgray] (0.4,0.2) rectangle (0.6,0.4);
        % \draw[black, very thick, fill=lightgray] (0,0.4) rectangle (0.2,0.6);
        % \draw[orange!0, very thick] (0,-0.1) rectangle (0,0.0);
        % \end{tikzpicture}} &
        % \cellcolor{orange!0}{
        % \begin{tikzpicture}[baseline=0.5em]
        % \draw[orange!0, very thick] (0,0.6) rectangle (0,0.7);
        % \draw[black, very thick, fill=lightgray] (0.4,0) rectangle (0.6,0.2);
        % \draw[black, very thick, fill=lightgray] (0,0.2) rectangle (0.2,0.4);
        % \draw[black, very thick, fill=lightgray] (0.2,0.2) rectangle (0.4,0.4);
        % \draw[black, very thick, fill=lightgray] (0.2,0.4) rectangle (0.4,0.6);
        % \draw[orange!0, very thick] (0,-0.1) rectangle (0,0.0);
        % \end{tikzpicture}}
        \\ \hline

        \begin{tabular}{c}
             $[[2(d-1)^2+2, 2, d]]$ \\
             \\
              \begin{tabular}{c|c}
                 \cellcolor{black!10}{\hspace{1em}$\bm{R_1}$\hspace{1em}} & 
        \cellcolor{black!10}{\hspace{1em}$\bm{R_2}$\hspace{1em}} \\
                \hline
                \cellcolor{orange!0}{\begin{tikzpicture}[baseline=0.5em]
                \draw[orange!0, very thick] (0,0.6) rectangle (0,0.7);
                \draw[black, very thick, fill=lightgray] (0.2,0) rectangle (0.4,0.2);
                \draw[black, very thick, fill=lightgray] (0.2,0.2) rectangle (0.4,0.4);
                \draw[black, very thick, fill=lightgray] (0.4,0.2) rectangle (0.6,0.4);
                \draw[black, very thick, fill=lightgray] (0,0.4) rectangle (0.2,0.6);
                \draw[orange!0, very thick] (0,-0.1) rectangle (0,0.0);
                \end{tikzpicture}} &
                \cellcolor{orange!0}{
                \begin{tikzpicture}[baseline=0.5em]
                \draw[orange!0, very thick] (0,0.6) rectangle (0,0.7);
                \draw[black, very thick, fill=lightgray] (0.4,0) rectangle (0.6,0.2);
                \draw[black, very thick, fill=lightgray] (0,0.2) rectangle (0.2,0.4);
                \draw[black, very thick, fill=lightgray] (0.2,0.2) rectangle (0.4,0.4);
                \draw[black, very thick, fill=lightgray] (0.2,0.4) rectangle (0.4,0.6);
                \draw[orange!0, very thick] (0,-0.1) rectangle (0,0.0);
                \end{tikzpicture}}\\
              \end{tabular}
        \end{tabular} & 
        \cellcolor{orange!0}{\hspace{.5em}
        \begin{tabular}{c}
        \\[-.7em]
        \begin{tabular}{c | c | c}
             \cellcolor{black!10}{\hspace{.3em} $\bm{d}$ \hspace{.3em}} & \cellcolor{black!10}{\hspace{.3em} $\bm{d_c}$ \hspace{.3em}} & \cellcolor{black!10}{\hspace{.3em} $\bm{v_\infty}$ \hspace{.3em}} \\
             \hline
             5 & 8 & 0.47\\
             6 & 10 & 0.38\\
             7 & 19 & 0.51\\
             \cellcolor{orange!50}{9} & \cellcolor{orange!50}{32} & \cellcolor{orange!50}{0.49}\\
             \cellcolor{orange!15}{10} & \cellcolor{orange!15}{30} & \cellcolor{orange!15}{0.37}\\
             \cellcolor{orange!15}{11} & \cellcolor{orange!15}{36} & \cellcolor{orange!15}{0.36}\\
             \cellcolor{orange!35}{13} & \cellcolor{orange!35}{60} & \cellcolor{orange!35}{0.41}\\
             \cellcolor{orange!35}{14} & \cellcolor{orange!35}{72} & \cellcolor{orange!35}{0.42}\\
             \cellcolor{orange!15}{15} & \cellcolor{orange!15}{67} & \cellcolor{orange!15}{0.34}\\
             \cellcolor{orange!35}{17} & \cellcolor{orange!35}{104} & \cellcolor{orange!35}{0.40}\\
             18 & 74 & 0.26\\
             19 & 76 & 0.23\\
             21 & 80 & 0.20\\
             22 & 82 & 0.19 \\
             23 & 264 & 0.27
        \end{tabular}
        \\
        \\[-.7em]
        \end{tabular}
        \hspace{.5em}
        } 
        %& 
        % \cellcolor{orange!0}{\begin{tikzpicture}[baseline=0.5em]
        % \draw[orange!0, very thick] (0,0.6) rectangle (0,0.7);
        % \draw[black, very thick, fill=lightgray] (0.2,0) rectangle (0.4,0.2);
        % \draw[black, very thick, fill=lightgray] (0.2,0.2) rectangle (0.4,0.4);
        % \draw[black, very thick, fill=lightgray] (0.4,0.2) rectangle (0.6,0.4);
        % \draw[black, very thick, fill=lightgray] (0,0.4) rectangle (0.2,0.6);
        % \draw[orange!0, very thick] (0,-0.1) rectangle (0,0.0);
        % \end{tikzpicture}} &
        % \cellcolor{orange!0}{
        % \begin{tikzpicture}[baseline=0.5em]
        % \draw[orange!0, very thick] (0,0.6) rectangle (0,0.7);
        % \draw[black, very thick, fill=lightgray] (0.4,0) rectangle (0.6,0.2);
        % \draw[black, very thick, fill=lightgray] (0,0.2) rectangle (0.2,0.4);
        % \draw[black, very thick, fill=lightgray] (0.2,0.2) rectangle (0.4,0.4);
        % \draw[black, very thick, fill=lightgray] (0.2,0.4) rectangle (0.4,0.6);
        % \draw[orange!0, very thick] (0,-0.1) rectangle (0,0.0);
        % \end{tikzpicture}}
        \\ \hline
        
    \end{tabular}
\end{table}

\begin{figure}[h!]
    \centering
    \includegraphics[width=\linewidth]{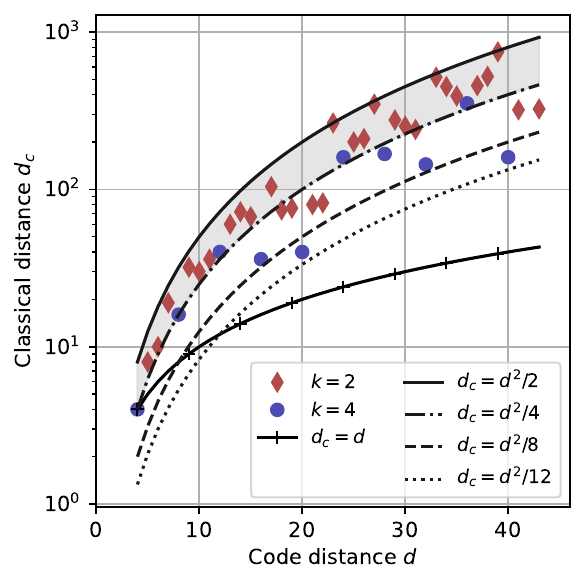}
    \caption{Classical distance as a function of the quantum distance for the codes in \cref{tab:Romanesco codes 2d plane extended} and additional ones at larger distances. For all points up to $d=17$ we used both BP+OSD-CS and an integer linear program decoder to compute the code distance.}
    \label{fig:dc_vs_d}
\end{figure}

It is more difficult to define code families on the two-dimensional plane than on the cylinder and torus because the classical distance is less predictable. In \cref{fig:dc_vs_d} we plot the classical distance $d_c$ as a function of the quantum code distance $d$ for all the codes in \cref{tab:Romanesco codes 2d plane extended} and some additional ones at larger distances. We also plotted lines proportional to $d^2$ to help identify families.

Despite not having defined clear families, we nonetheless emphasize that \cref{fig:dc_vs_d} clearly shows that the classical distance does grow with $2d^2>n$, similar to what is observed for the codes on the torus and the cylinder. Particularly, we can see that for $k=4$ the codes which have $d$ equal to an integer multiple of 12 have a classical distance larger than $n/8$. However, we remark that $d_c$ does not follow the same scalings observed on the torus and the cylinder (see \cref{tab:Romanesco codes rectangular,tab:Romanesco codes cylinder}) where we instead have $d_c \sim 3n/16$ for all codes with lattice dimensions that are integer multiples of 4. It is possible that a different way of opening the lateral boundaries could result in a different scaling for the classical distance $d_c$.

Overall, more optimization could be done for the boundary, so that we achieve better code parameters.

\subsection{Logical representatives}

Here we investigate the minimum-weight mixed logical representatives on the two-dimensional plane. We sample the representatives like we did on the torus (see \cref{app:logical representatives}). The distribution is shown in \cref{tab:open BC mixed logicals} for the small to moderate size codes in \cref{tab:Romanesco codes 2d plane extended}. Importantly, for codes with $k=4$, the lattice dimensions of each sub-lattice is odd and therefore we now have representatives with one Pauli X operator. Those will impact the effective distance of the code depending on the noise bias. As noted in the main text, the $[[244, 4, 12]]$ code has a classical distance 40 but the minimum weight of logical representatives with a single X Pauli operator is 40 and therefore, when looking at the memory experiment with code-capacity noise, we see that the code distance at large noise bias is closer to 22 than 40. The distribution in \cref{tab:open BC mixed logicals} also depends on how we truncate the stabilizers at the boundary and it might be possible to improve the performance of the codes using a different approach.

\begin{table}[h!]
    \centering
    \caption{Distance $d_s$ of the lowest-weight logical representative with $s$ non-Z Pauli operators for some of the codes in \cref{tab:Romanesco codes 2d plane extended}. The saturation is proportional to the drop in distance relative to $s=0$. The codes that are highlighted do not have a distance drop with one bit-flip error.}
    \label{tab:open BC mixed logicals}
    \begin{tabular}{c c c c c c c c c c c}
        %\hline
        & \multicolumn{10}{c}{\cellcolor{black!10}{$\bm{s}$}} \\
        %\hline
        \cellcolor{black!10}{$\bm{[[n,k,d]]}$} & \cellcolor{black!10}{$\bm{0}$} & \cellcolor{black!10}{$\bm{1}$} & \cellcolor{black!10}{$\bm{2}$} & \cellcolor{black!10}{$\bm{3}$} & \cellcolor{black!10}{$\bm{4}$} & \cellcolor{black!10}{$\bm{5}$} & \cellcolor{black!10}{$\bm{6}$} & \cellcolor{black!10}{$\bm{7}$} & \cellcolor{black!10}{$\bm{8}$} & \cellcolor{black!10}{$\bm{9}$}\\ %\hline
        
        \cellcolor{black!20}{$[[20,4,4]]$} & \cellcolor{orange!10}{4} & \cellcolor{orange!10}{4} & \cellcolor{orange!10}{4} & \cellcolor{black!75} & \cellcolor{black!75} & \cellcolor{black!75} & \cellcolor{black!75} & \cellcolor{black!75} & \cellcolor{black!75}  & \cellcolor{black!75} \\ %\hline

        \cellcolor{black!10}{$[[34,2,5]]$} & \cellcolor{orange!10}{8} & \cellcolor{orange!90}{5} & \cellcolor{orange!90}{5} & \cellcolor{orange!90}{5} & \cellcolor{black!75} & \cellcolor{black!75} & \cellcolor{black!75} & \cellcolor{black!75} & \cellcolor{black!75} & \cellcolor{black!75} \\ %\hline

        \cellcolor{black!20}{$[[52,2,6]]$} & \cellcolor{orange!10}{10} & \cellcolor{orange!10}{10} & \cellcolor{orange!90}{6} & \cellcolor{orange!90}{6} & \cellcolor{black!75} & \cellcolor{black!75} & \cellcolor{black!75} & \cellcolor{black!75} & \cellcolor{black!75} & \cellcolor{black!75} \\ %\hline

        \cellcolor{black!10}{$[[74,2,7]]$} & \cellcolor{orange!10}{19} & \cellcolor{orange!40}{10} & \cellcolor{orange!80}{9} & \cellcolor{orange!90}{7} & \cellcolor{orange!90}{7} & \cellcolor{black!75} & \cellcolor{black!75} & \cellcolor{black!75} & \cellcolor{black!75} & \cellcolor{black!75} \\ %\hline

        \cellcolor{black!10}{$[[100,4,8]]$} & \cellcolor{orange!10}{16} & \cellcolor{orange!80}{10} & \cellcolor{orange!90}{8} & \cellcolor{orange!80}{10} & \cellcolor{orange!90}{8} & \cellcolor{black!75} & \cellcolor{black!75} & \cellcolor{black!75} & \cellcolor{black!75} & \cellcolor{black!75}  \\ %\hline

        \cellcolor{black!10}{$[[130,2,9]]$} & \cellcolor{orange!10}{32} & \cellcolor{orange!40}{19} & \cellcolor{orange!80}{11} & \cellcolor{orange!90}{9} & \cellcolor{orange!90}{9} & \cellcolor{orange!90}{9} & \cellcolor{black!75} & \cellcolor{black!75} & \cellcolor{black!75} & \cellcolor{black!75} \\ %\hline

        \cellcolor{black!10}{$[[164,2,10]]$} & \cellcolor{orange!10}{30} & \cellcolor{orange!40}{18} & \cellcolor{orange!70}{14} & \cellcolor{orange!80}{18} & \cellcolor{orange!90}{12} & \cellcolor{orange!90}{10} & \cellcolor{black!75} & \cellcolor{black!75} & \cellcolor{black!75} & \cellcolor{black!75} \\ %\hline
        
        \cellcolor{black!20}{$[[202,2,11]]$} & \cellcolor{orange!10}{36} & \cellcolor{orange!40}{41} & \cellcolor{orange!70}{17} & \cellcolor{orange!80}{14} & \cellcolor{orange!80}{13} & \cellcolor{orange!90}{11} & \cellcolor{orange!90}{11} & \cellcolor{black!75} & \cellcolor{black!75} & \cellcolor{black!75} \\ %\hline

        \cellcolor{black!10}{$[[244,4,12]]$} & \cellcolor{orange!10}{40} & \cellcolor{orange!40}{22} & \cellcolor{orange!70}{16} & \cellcolor{orange!80}{14} & \cellcolor{orange!90}{12} & \cellcolor{orange!90}{12} & \cellcolor{orange!90}{12} & \cellcolor{black!75} & \cellcolor{black!75} & \cellcolor{black!75} \\ %\hline

        \cellcolor{black!20}{$[[290,2,13]]$} & \cellcolor{orange!10}{60} & \cellcolor{orange!40}{$\varnothing$} & \cellcolor{orange!70}{23} & \cellcolor{orange!80}{17} & \cellcolor{orange!80}{17} & \cellcolor{orange!80}{13} & \cellcolor{orange!90}{13} & \cellcolor{orange!90}{13} & \cellcolor{black!75} & \cellcolor{black!75} \\ %\hline

        \cellcolor{black!10}{$[[340,2,14]]$} & \cellcolor{orange!10}{72} & \cellcolor{orange!40}{48} & \cellcolor{orange!70}{24} & \cellcolor{orange!80}{22} & \cellcolor{orange!80}{20} & \cellcolor{orange!70}{26} & \cellcolor{orange!90}{$\varnothing$} & \cellcolor{orange!90}{14} & \cellcolor{black!75} & \cellcolor{black!75} \\ %\hline

        \cellcolor{black!10}{$[[394,2,15]]$} & \cellcolor{orange!10}{67} & \cellcolor{orange!40}{50} & \cellcolor{orange!70}{29} & \cellcolor{orange!80}{23} & \cellcolor{orange!80}{21} & \cellcolor{orange!70}{33} & \cellcolor{orange!90}{$\varnothing$} & \cellcolor{orange!90}{15} & \cellcolor{orange!90}{15} & \cellcolor{black!75} \\ %\hline

        \cellcolor{black!20}{$[[452,4,16]]$} & \cellcolor{orange!10}{36} & \cellcolor{orange!40}{70} & \cellcolor{orange!70}{28} & \cellcolor{orange!80}{22} & \cellcolor{orange!80}{20} & \cellcolor{orange!80}{22} & \cellcolor{orange!90}{16} & \cellcolor{orange!90}{16} & \cellcolor{orange!90}{16} & \cellcolor{black!75} \\ %\hline

        \cellcolor{black!10}{$[[514,2,17]]$} & \cellcolor{orange!10}{104} & \cellcolor{orange!40}{69} & \cellcolor{orange!70}{35} & \cellcolor{orange!80}{29} & \cellcolor{orange!80}{27} & \cellcolor{orange!80}{25} & \cellcolor{orange!90}{$\varnothing$} & \cellcolor{orange!90}{17} & \cellcolor{orange!90}{17} & \cellcolor{orange!90}{17} \\ %\hline
        
    \end{tabular}
\end{table}

\section{Numerical memory experiments}\label{app:sims}

In this appendix, we detail how we generate the logical performance data. We run Monte Carlo simulations with code-capacity noise. We solve the decoding problem as described in \cref{app:rotating decoding problem} with the CSS code parity check matrix and rotated noise channels (i.e. a Hadamard rotation is applied on each gray qubit). Importantly, we need to define a logical basis so that we can compute the X and Z logical error rates. We use \cref{algo:logical basis} to define this basis. We use an infinitely-biased noise model with noise bias $\eta = \infty$ and depolarizing rate $p$ so that $p_X \to 0$ and $p_Z = p$. This algorithm uses the parity check matrix of the CSS code along with the rotated noise channels, in the same way we run the code-capacity simulations. With this basis, we expect no logical X error in the infinite bias limit. We generate random errors on both the X and Z components of the code using the probabilities in the noise channels. We multiply the parity check matrix with these errors modulo 2 to get the syndromes. We find the corrections by decoding those syndromes and apply them on the error vectors. We then determine if there is any logical error by applying the logical operators on the error vectors modulo 2. The X and Z logical error rates are the number of X and Z logical errors divided by the number of shots.

\begin{algorithm}[h!]
    \caption{
    We solve a modified decoding problem to obtain a logical basis for any CSS code with parity check matrices $H_X$ and $H_Z$ in order to find minimum-weight logical representatives with different numbers of X Pauli operators. The roles of $H_X$ and $H_Z$ can be swapped. Here we use the BP+OSD-CS decoder with the same parameters as in \cref{algo:code distance}.}\label{algo:logical basis}
    \begin{algorithmic}
        \Procedure{Logical basis}{$H_X$, $H_Z$, $[p_{Xi}]$, $[p_{Zi}]$}
        \State \textbf{input: } X parity check matrix $H_X$, Z parity check matrix $H_Z$, X depolarizing rate per qubit $[p_{Xi}]$, Z depolarizing rate per qubit $[p_{Zi}]$
        \State \textbf{output: } Orthogonal basis of X and Z logicals
        \State $n = \mathrm{ncol}(H_X)$
        \State $k = n -\mathrm{rk}_{\mathrm{mod}2}(H_X)-\mathrm{rk}_{\mathrm{mod}2}(H_Z)$
        \State $[L_X] \gets$ list of X logicals
        \State $[L_Z] \gets$ list of Z logicals
        \While{$\mathrm{len}([L_Z])\neq k$}
        \State // \textit{Get test X logicals}
        \State $[\tilde{L}_X] \gets$ list of X test logicals
        \For{$\tilde{L}_X \in \mathrm{ker}_{\mathrm{mod}2}(H_Z)$}
        \State $\tilde{H}_X = \mathrm{vstack}(H_X, \tilde{L}_X)$
        \State // \textit{check it is not in the rowspace of $H_X$}
        \If{$\mathrm{rk}_{\mathrm{mod}2}(\tilde{H}_X)\neq \mathrm{rk}_{\mathrm{mod}2}(H_X)$}
        \State add $\tilde{L}_X$ to $[\tilde{L}_X]$
        \EndIf
        \EndFor
        \State // \textit{Find low-weight Z logicals}
        \State $[\tilde{L}_Z] \gets$ list of test Z logicals
        \State // \textit{use random subset of $[\tilde{L}_X]$}
        \For{$\tilde{L}_X \in [\tilde{L}_X]$}
        \State // \textit{set up the decoder}
        \State $D_X = \mathrm{BPOSD}(\mathrm{vstack}(H_X,\tilde{L}_X), [p_{Xi}])$
        \State $s_X = \mathrm{vstack}(\vec{0}_{\mathrm{nrow}(H_X)}, 1)$ // syndrome
        \State $\tilde{L}_Z = D_X(s_X)$ // \textit{decode to get Z logical}
        \State add $L_Z$ to $[\tilde{L}_Z]$
        \EndFor
        \State sort $[\tilde{L}_Z]$ by length
        \For{$L_Z \in [\tilde{L}_Z]$}
        \State //\textit{check Z logical is linearly independent}
        \State $\tilde{H}_Z = \mathrm{vstack}(H_Z, L_Z)$
        \State $q = 0$ // \textit{count the number of new Z logicals}
        \If{$\mathrm{rk}_{\mathrm{mod}2}(\tilde{H}_Z)\neq \mathrm{rk}_{\mathrm{mod}2}(H_Z)$}
        \State $H_Z=\tilde{H}_Z$
        \State add $L_Z$ to $[L_Z]$
        \State $q = q + 1$
        \EndIf
        \State \textbf{if} $\mathrm{len}([L_Z])=k$ \textbf{then} stop
        \EndFor
        \State // \textit{Find X logicals to complete the basis}
        \State $D_Z = \mathrm{BPOSD}(H_Z, [p_{Zi}])$ // set up the decoder
        \ForAll{$ 0 \leq i <q$}
        \State $s_Z = \mathrm{vstack}(\vec{0}_{\mathrm{nrow}(H_Z)-q+i}, 1, \vec{0}_{q-i-1})$
        \State $L_X = D_Z(s_Z)$ // \textit{decode to get X logical}
        \State $H_X \to \mathrm{vstack}(H_X, L_X)$
        \State add $L_X$ to $[L_X]$
        \EndFor
        \EndWhile
        \Return $[L_X]$, $[L_Z]$
        \EndProcedure
    \end{algorithmic}
\end{algorithm}

\section{Modified BP+OSD decoder in the large bias regime}\label{app: custom decoder}

In order to improve the performance of BP+OSD decoder in the strong bias regime, we solve multiple decoding problems in parallel with BP+OSD and keep the correction $C = (C_Z | C_X)$ (where $C_{Z/X}$ is the X/Z correction) with the lowest weight $\sum_{q \in C_Z}\log(p_Z^{-1}-1)+\sum_{q \in C_X}\log(p_X^{-1}-1)$. We solve a total of $n+2$ decoding problems. Recall that we simulate the non-CSS Romanesco codes by simulating the CSS version with a rotated noise model. First, we split the X and Z part of the syndrome and use the BP+OSD-CS decoder with the $H_X$ and $H_Z$ parity check matrix of the CSS version of the code. Second, we solve a classical decoding problem with BP+OSD-E by eliminating the half of each stabilizer generator that detects X errors in the non-CSS Romanesco codes, thereby assuming only Z errors caused the syndrome. As in~\cref{fig:xyz ca rules}(a), this amounts to removing the gray qubits from every Z check and the black qubits from every X check in the CSS code. The output is the correction that we expect at infinite bias. To account for one bit-flip errors we solve this classical decoding problem $n$ other times: we artificially apply a bit-flip to one of the data qubits and modify the syndrome accordingly. The bit-flip is subsequently added to the correction that the decoder returns.

\section{Decoder comparison} \label{app: decoder comparison}

\begin{figure}[ht!]
    \centering
    \includegraphics[width=.85\linewidth]{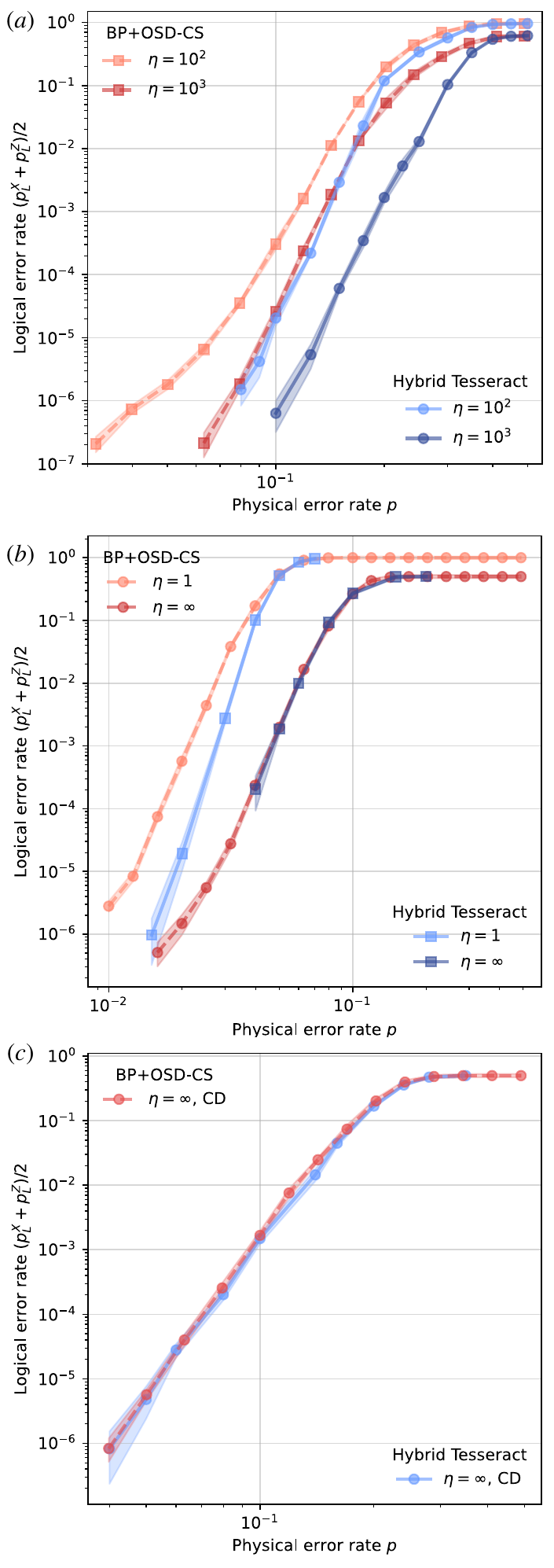}
    \caption{Logical error rate of (a) $[[288, 12, 12]]_\clubsuit^{\protect \torus}$, (b) $[[288, 12, 18]]_{\rm BB}^{\protect \torus}$ and (c) Clifford-deformed $[[288, 12, 18]]_{\rm BB}^{\protect \torus}$ as a function of the physical error rate using the BP+OSD-CS decoder or the hybrid tesseract decoder. The shaded regions identify the 95\% confidence intervals.}
    \label{fig:bposd vs tesseract}
\end{figure}

\begin{figure}[ht!]
    \centering
    \includegraphics[width=.85\linewidth]{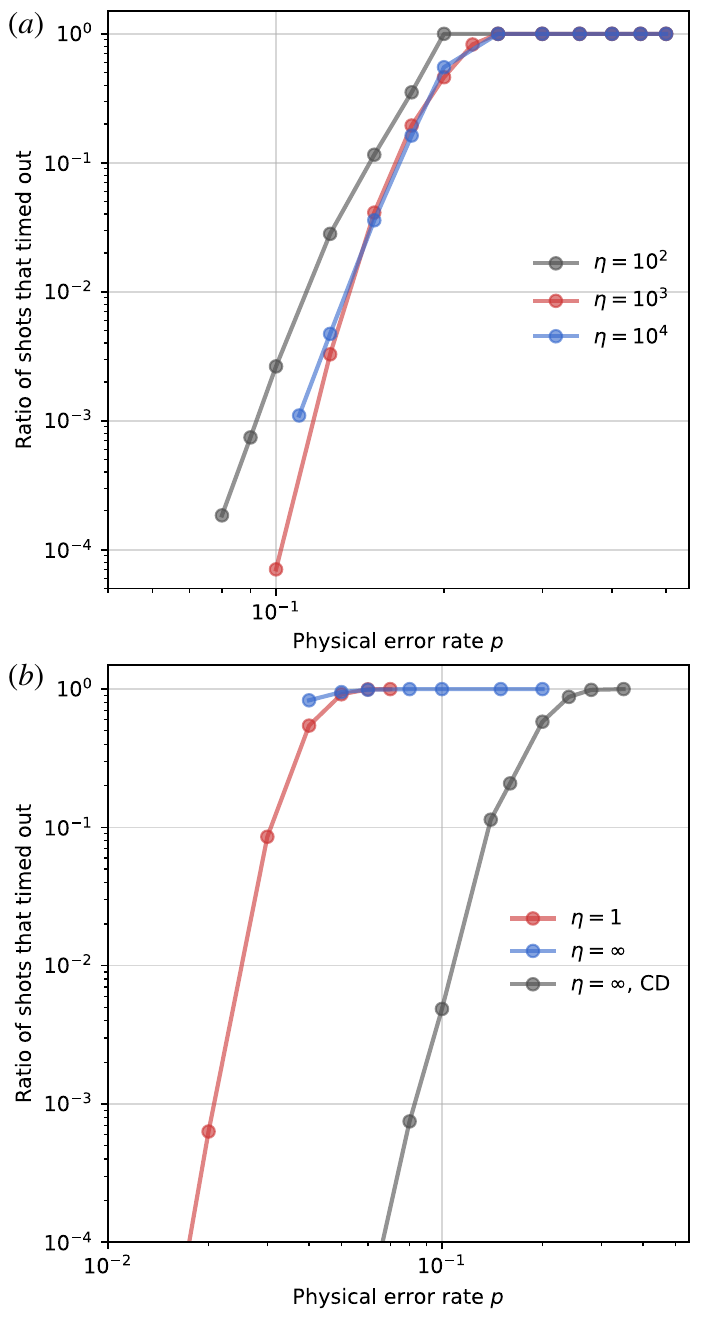}
    \caption{The fraction of shots where the tesseract decoder ran for more than a minute for (a) $[[288, 12, 12]]_\clubsuit^{\protect \torus}$ and (b) t$[[288, 12, 18]]_{\rm BB}^{\protect \torus}$. Here ``CD'' stands for Clifford-deformed.}
    \label{fig:timeout ratios}
\end{figure}

We studied the logical error rates for code-capacity noise with BP+OSD-CS and a hybrid tesseract decoder. This hybrid decoder attempts to run the tesseract decoder, but any shot that is not decoded within one minute is decoded with a fallback decoder. For the non-Romanesco codes we use BP+OSD-CS with 1000 BP iterations and OSD order 50. For Romanesco codes we fall back to the modified decoder in \cref{app: custom decoder}. For the tesseract decoder we used beam climbing and beam factor 40.

We found significant improvements from the tesseract decoder for Romanesco codes under biased noise. In~\cref{fig:bposd vs tesseract} we compare for (a) the $[[288, 12, 12]]_\clubsuit^{\protect \torus}$ Romanesco codes, (b) the $[[288, 12, 18]]_{\rm BB}^{\protect \torus}$, and (c) the Clifford-deformed $[[288, 12, 18]]_{\rm BB}^{\protect \torus}$ code. For the Romanesco code we observe both lower logical error rate and improved slope for smaller values of $p$. The scaling with $p$ we find with the BP+OSD-CS decoder is worse than we expect based on the distance of the code. With the tesseract decoder we recover the expected scaling. 

For the bivariate bicycle code $[[288, 12, 16]]_{\rm BB}^{\torus}$, we observed significant improvement due to the tesseract decoder only for the case of no bias $\eta = 1$~\cref{fig:bposd vs tesseract}(b). For infinite bias, which we simulated with the tesseract decoder by setting $\eta = 10^9$, we observed that the hybrid tesseract decoder matches BP+OSD-CS. This is because the tesseract decoder did not run successfully within one minute for most of the shots as in~\cref{fig:timeout ratios}(b). 

For the Clifford-deformed $[[288, 12, 16]]_{\rm BB}^{\torus}$ code with infinite bias, the tesseract decoder ran within a minute, but nevertheless the logical error rate was not improved on average. 

For every code, we find that the tesseract decoder struggles to run within one minute as $p$ increases, or as the syndrome density increases. As in~\cref{fig:timeout ratios}, for lower values of $p$ the decoder is eventually able to run on nearly all shots. However, the $[[288, 12, 18]]_{\rm BB}^{\protect \torus}$ bivariate bicycle code with infinite bias seems to be much more challenging for the decoder than the other cases.

\newpage

\bibliography{refs}

\end{document}